\documentclass[]{article}

\usepackage{amsfonts}
\usepackage{amsthm}
\usepackage{url}
\usepackage{amsmath}
\usepackage{graphics}
\usepackage{color}

\newtheorem{defn}{Definition}
\newtheorem{theorem}{Theorem}
\newtheorem{prop}{Proposition}
\newtheorem{lemma}{Lemma}
\newtheorem{cor}{Corollary}

\newcommand{\Var}{\ensuremath{\mathrm{Var}}}
\newcommand{\Cov}{\ensuremath{\mathrm{Cov}}}
\newcommand{\markov}{\ensuremath{\leftrightarrow}}
\newcommand{\diag}{\ensuremath{\mathrm{diag}}}
\newcommand{\bmu}{\mbox{\boldmath\ensuremath{\mu}}}
\newcommand{\m}[1]{\mbox{\boldmath\ensuremath{#1}}}
\newcommand{\ve}[1]{\mathbf{#1}}
\newcommand{\SNR}{\ensuremath{\mathsf{SNR}}}

\begin{document}

\title{Rate Region of the Quadratic Gaussian \\
Two-Encoder Source-Coding Problem}

\author{Aaron B.\ Wagner\thanks{School of Electrical and Computer Engineering, 
Cornell University. Email: \texttt{wagner@ece.cornell.edu}.},
Saurabha Tavildar\thanks{QUALCOMM Flarion Technologies.
Email: \texttt{tavildar@gmail.com}.},
and Pramod Viswanath\thanks{Department of Electrical and Computer
Engineering and Coordinated Science Laboratory, 
University of Illinois at Urbana-Champaign. Email: \texttt{pramodv@uiuc.edu}.
Much of this work was conducted while the first two authors were with the
University of Illinois.}
}

\date{February 8, 2008}

\maketitle

\begin{abstract}
We determine the rate region of the quadratic Gaussian
two-encoder source-coding problem. 
This rate region is achieved by a simple architecture
that separates the analog and digital aspects of
the compression. Furthermore, this architecture requires
higher rates to send a Gaussian source than it does to
send any other source with the same covariance.
Our techniques can also be used to
determine the sum rate of some generalizations of
this classical problem.
Our approach involves coupling
the problem to a quadratic Gaussian ``CEO problem.''
\end{abstract}


\section{Introduction}

\begin{figure}
\begin{center}
\scalebox{1.1}{\input{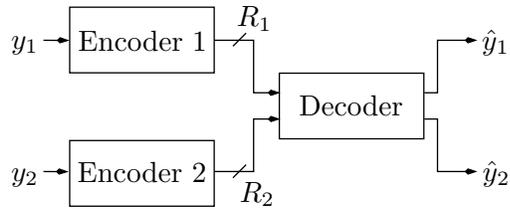}}
\end{center}
\caption{The two-encoder source-coding problem.}
\label{setup}
\end{figure}

This paper addresses the quadratic Gaussian two-encoder
source-coding problem, the setup for which is 
depicted in Fig.~\ref{setup}. 
Two encoders observe different components of a memoryless, Gaussian,
vector-valued source. The encoders, without cooperating,
compress their observations and
send messages
to a single decoder over rate-constrained, noiseless
channels. The decoder attempts to reproduce both observations,
subject to separate constraints on the time-averaged expected
squared
error of the two estimates. We seek to determine the set of
rate pairs $(R_1,R_2)$ that allow us to meet a given pair of
target distortions. We call this set the
\emph{rate region}. Of course, this problem
can also be formulated for general sources and
distortion measures. Our focus on the 
quadratic Gaussian case is motivated by its fundamental nature,
its importance in applications, and its well-known extremal
properties.

This problem is naturally viewed as a quadratic Gaussian version of
Slepian and Wolf's problem~\cite{Slepian:SW}.
Slepian and Wolf studied the problem in which
the source is discrete and the decoder
must reproduce it with negligible probability of error.
Their celebrated result
characterizes the rate region for this setup. One
consequence of this characterization is that permitting the
encoders to cooperate or view each other's observations does
not offer any advantage, at least as far as the sum rate
is concerned.

\begin{figure}
\begin{center}
\scalebox{1.1}{\input{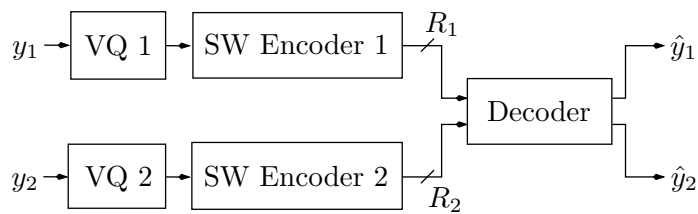}}
\end{center}
\caption{A natural architecture that separates the analog and
digital aspects of the compression.}
\label{natsep}
\end{figure}

There is a natural way to harness Slepian and Wolf's result
in the Gaussian setting. Each encoder first vector quantizes (VQs)
its observation using a Gaussian test channel as in single-encoder
rate-distortion theory.
This results in two correlated digital messages, which are 
suitable for compression via Slepian-Wolf encoding. The
decoder decodes the quantized values and estimates
the source by computing a conditional expectation. This 
approach separates the analog and digital aspects of the
compression, as shown in Fig.~\ref{natsep}.

Our main result is an explicit characterization
of the rate region for this problem. This result has three
notable consequences:
\begin{enumerate}
\item[(\emph{i})] The architecture depicted in
Fig.~\ref{natsep} is optimal.
\item[(\emph{ii})] This architecture requires higher rates to
send a Gaussian source than it does to send any other source with
the same covariance. In 
particular, a Gaussian source has the smallest rate region for
a given covariance.
\item[(\emph{iii})] Unlike in the Slepian-Wolf problem, here
decentralized encoding requires a strictly higher rate compared
to centralized encoding.
\end{enumerate}

The problem of determining the rate region for this setup has 
been open for some time~\cite{Berger:MTSC}. Early 
work~\cite{Berger:MTSC,Tung:PHD} used
the architecture described above to prove an inner bound.
Zamir and Berger~\cite{Zamir:Resolution} showed that this inner bound is 
asymptotically tight
in the low-distortion regime, even if the source is not Gaussian.
Oohama~\cite{Oohama:Gaussian:MTSC} determined the rate
region for the problem in which only one of
the two distortion constraints is present. By
interpreting this problem as a relaxation of
the original problem, he obtained an outer bound
for the latter. He showed
that this outer bound, when combined with the
inner bound, determines a portion
of the boundary of the rate region. As a result
of his work, showing that the inner bound is tight
in the sum rate suffices to complete the characterization
of the rate region. This is shown in the present paper.

Our approach is to lower bound the sum rate of a given
code in two different ways. The first way amounts to 
considering the rate required by a hypothetical centralized 
encoder that achieves the same error covariance matrix
as the code. The second way is to establish a 
connection between this problem and the quadratic
Gaussian ``CEO problem,'' for which the rate region
is already known.
For some codes, the cooperative bound may be tighter. For 
others, the CEO bound may be tighter. Taking the 
maximum of the two lower bounds yields a composite lower
bound that is sufficiently strong to prove the desired result.

The next section contains a precise formulation of the problem
and a statement of our main result, Theorem~\ref{main:theorem}.
In Section~\ref{achieve:section}, we describe
the separation-based compression
architecture. There we also discuss the worst-case property of the Gaussian
distribution. 
We provide the necessary background on the CEO problem
and some other preliminaries to the converse proof
in Section~\ref{CEO:section}.
The converse proof itself is contained 
in Section~\ref{proof:section}. In Section~\ref{remote:section}, we
show how the converse proof technique can be used to
determine the rate region for a more general version of the
problem in which the decoder aims to reproduce certain
linear combinations of the source components. In Section~\ref{many:section},
we show how the proof technique can be extended to
handle the case of more than two sources, if a certain
symmetry condition holds.
Section~\ref{remarks:section} contains some concluding 
remarks.

We use the following notation. Boldface, lower case
letters ($\bmu$) denote vectors, while boldface, upper case
letters (\m{D}) denote matrices. Lightface
letters ($\rho$, $R$) denote scalars. Whether a variable is
deterministic or random should be clear from the context.

\section{Problem Formulation and Main Result}
\label{prelim:sec}

Let $\{(y_1^n(i),y_2^n(i))\}_{i = 1}^n$ be a sequence of
independent and identically distributed (i.i.d.) Gaussian
zero-mean random vectors. Let
\begin{equation}
\m{K}_y = 
\left[
\begin{array}{cc}
1 & \rho \\
\rho & 1
\end{array}
\right]
\end{equation}
denote the covariance matrix of $(y_1^n(1),y_2^n(1))$.
We use $y_1^n$ to denote 
\begin{equation*}
\{y_1^n(i)\}_{i = 1}^n,
\end{equation*}
$y_1^n(j:k)$ to denote
\begin{equation*}
\{y_1^n(i)\}_{i = j}^k,
\end{equation*}
$\ve{y}^n(i)$ to denote $(y_1^n(i),y_2^n(i))$,
$\ve{y}^n$ to denote $\{(y_1^n(i),y_2^n(i))\}_{i = 1}^n$, etc.
Analogous notation will be used for other vectors
that appear later.

The first encoder observes $y_1^n$, then
sends a message to the decoder using a mapping
\begin{equation*}
f_1^{(n)} : \mathbb{R}^n \mapsto \left\{1,\ldots,M_1^{(n)}\right\}.
\end{equation*}
The second encoder operates analogously. The decoder 
uses the received messages to estimate both 
$y^n_1$ and $y^n_2$ using mappings
\begin{equation*}
\varphi_j^{(n)} : \left\{1,\ldots,M_1^{(n)}\right\} \times 
   \left\{1,\ldots,M_2^{(n)}\right\} \mapsto
   \mathbb{R}^n  \quad j = 1,2.
\end{equation*}

\begin{defn}[Quadratic Gaussian Two-Encoder Source Coding Problem]
A rate-distortion vector $(R_1,R_2,d_1,d_2)$ 
is \emph{strict-sense achievable} if there exists
a block length $n$, encoders $f_1^{(n)}$ and $f_2^{(n)}$, and a decoder
$(\varphi^{(n)}_1,\varphi^{(n)}_2)$ such that\footnote{All logarithms
in this paper are base two.}
\begin{equation}
\label{feasible}
\begin{split}
R_j & \ge \frac{1}{n} \log M_j^{(n)} \ \text{for all $j$ in \{1,2\}, and} \\
d_j & \ge \frac{1}{n} \sum_{i = 1}^n 
   E\left[\left(y^n_j(i) - \hat{y}^n_j(i)\right)^2\right] 
   \ \text{for all $j$ in \{1,2\}},
\end{split}
\end{equation}
where 
\begin{equation*}
\hat{y}_j^n = \varphi_j^{(n)}\left(f_1^{(n)}(y_1^n),f_2^{(n)}(y_2^n)\right) 
  \quad j \in \{1,2\}.
\end{equation*}
Let $\mathcal{RD}^\star$ denote the set of strict-sense achievable 
rate-distortion vectors. We define the set of \emph{achievable} 
rate-distortion vectors to be the closure, $\overline{\mathcal{RD}^\star}$,
of $\mathcal{RD}^\star$. Let 
\begin{equation*}
\mathcal{R}^\star(d_1,d_2) = \left\{(R_1,R_2) : (R_1,R_2,d_1,d_2) \in 
   \overline{\mathcal{RD}^\star}\right\}.
\end{equation*}
We call $\mathcal{R}^\star(\cdot,\cdot)$ the
\emph{rate region} for the problem. The \emph{(minimum) sum
rate} for a given distortion pair $(d_1,d_2)$ is defined to be
\begin{equation*}
\inf \{R_1 + R_2 : (R_1,R_2) \in \mathcal{R}^\star(d_1,d_2)\}.
\end{equation*}
\end{defn}

We note that there is no loss of generality in assuming
that $E[y_1^2] = E[y_2^2] = 1$, since the observations
and the estimates can be scaled to reduce the general
case to this one. By similar reasoning, we may assume that
$\rho \ge 0$, i.e., that the observations of the two encoders
are nonnegatively correlated. Since the two extreme cases $\rho = 0$
and $\rho = 1$ can be handled using existing techniques, 
we will assume throughout the remainder of the paper that $0 < \rho < 1$.

We now define three sets that will be used to describe
the rate region. Let
\begin{equation*}
\mathcal{R}_1^{\star}(d_1) = \left\{(R_1,R_2) : R_1 \ge 
   \frac{1}{2} \log^+ \left[ \frac{1}{d_1} 
   \left(1 - \rho^2 + \rho^2 2^{-2 R_2}\right)
   \right]\right\},
\end{equation*}
where $\log^+ x = \max(\log x, 0)$. Likewise, let
\begin{equation*}
\mathcal{R}_2^{\star}(d_2) = \left\{(R_1,R_2) : R_2 \ge 
   \frac{1}{2} \log^+ \left[ \frac{1}{d_2} 
   \left(1 - \rho^2 + \rho^2 2^{-2 R_1}\right)
   \right]\right\}.
\end{equation*}
Finally, let
\begin{equation*}
\mathcal{R}_{\mathrm{sum}}^\star(d_1,d_2) = \left\{(R_1,R_2) : R_1 + R_2 \ge
  \frac{1}{2} \log^+ \left[ \frac{(1-\rho^2) \; \beta(d_1,d_2)}{2 d_1 d_2}
   \right]\right\},
\end{equation*}
where
\begin{equation*}
\beta(d_1,d_2) = 1 + \sqrt{1 + \frac{4 \rho^2 d_1 d_2}{(1 - \rho^2)^2}}.
\end{equation*}

Later we will see that we can often interpret
the logarithm in the definition of
$\mathcal{R}_\mathrm{sum}^\star(\cdot,\cdot)$
as a mutual information
\begin{equation*}
 \frac{1}{2} \log \frac{|\m{K}_y|}{|\m{D}^*|},
\end{equation*}
where $\m{D}^*$ is the covariance matrix of the errors $(y_1 - \hat{y}_1,
y_2 - \hat{y}_2)$ in a sum-rate optimal code. Throughout the
paper
we assume that all distortion constraints ($d_1$ and $d_2$ in this
case) are positive.

\begin{theorem}
\label{main:theorem}
For the Gaussian two-encoder source-coding problem,
\begin{equation}
\mathcal{R}^\star(d_1,d_2) = \mathcal{R}_1^\star(d_1) \cap
   \mathcal{R}_2^\star(d_2) \cap \mathcal{R}_{\mathrm{sum}}^\star(d_1,d_2).
\end{equation}
\end{theorem}

\begin{figure}
\begin{center}
\scalebox{.60}{\input{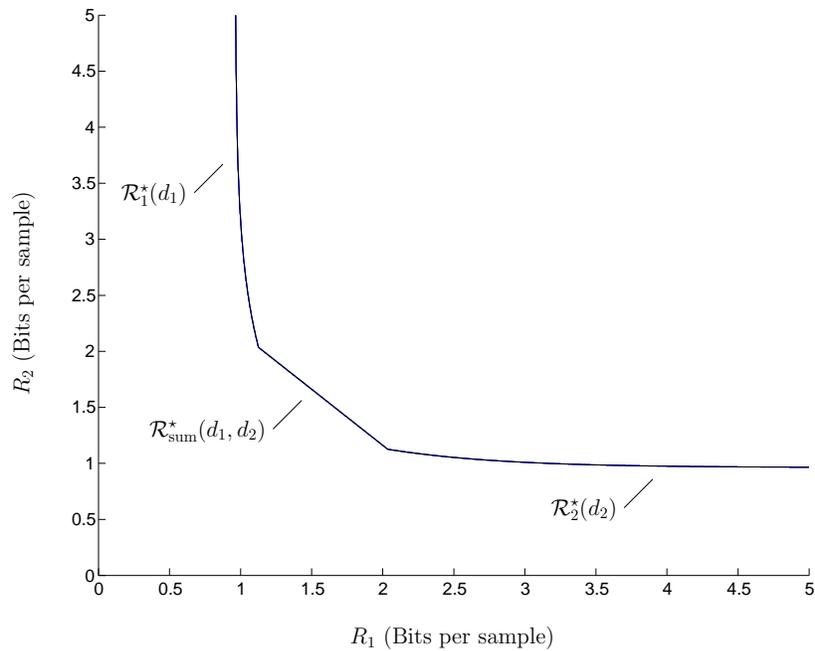}}
\end{center}
\caption{The rate region for $\rho = 0.9$ and $d_1 = d_2 = 0.05$.}
\label{region}
\end{figure}

An example of the rate region is shown in Fig.~\ref{region}.
The direct part of this result was previously known and
is discussed in the next
section. It was also previously known that the rate region was
contained in the set $\mathcal{R}_1^\star(d_1) \cap
\mathcal{R}_2^\star(d_2)$. Our contribution is a proof
that the rate region in contained in 
$\mathcal{R}_{\mathrm{sum}}^\star(d_1,d_2)$.
This is provided in Sections~\ref{CEO:section} 
and~\ref{proof:section}. Sections~\ref{remote:section} and~\ref{many:section} 
present some extensions of this result to problems with
more general distortion constraints and more than two
sources, respectively.

\section{Direct Part and Worst-Case Property}
\label{achieve:section}

Translating the architecture in Fig.~\ref{natsep} into an inner
bound on the rate region is a straightforward exercise in network
information theory. Since proofs of similar bounds are 
available~\cite{Berger:MTSC,Berger:CEO,Gastpar:WZ,Oohama:CEO:Region,
Viswanathan:CEO},
we provide only a high-level view of the proof here.

Let $\mathcal{U}(d_1,d_2)$ denote the set of real-valued
random variables $u_1$
and $u_2$ such that
\begin{enumerate}
\item[(\emph{i})] $u_1 \markov y_1 \markov y_2 \markov u_2$, meaning
that $u_1$, $y_1$, $y_2$, and $u_2$ form a Markov chain in this 
order\footnote{This condition is sometimes called the 
``long Markov chain''~\cite{Zamir:Resolution}.},
and
\item[(\emph{ii})]
$E[(y_j - E[y_j|\ve{u}])^2] \le d_j$ for $j \in \{1,2\}$.
\end{enumerate}
Then fix $\ve{u}$ in $\mathcal{U}(d_1,d_2)$ and a large integer
$n$. By the proof of the point-to-point rate-distortion theorem,
the first vector quantizer can send $I(y_1;u_1)$ bits per sample
to the first Slepian-Wolf encoder that conveys a string $u_1^n$ 
that is jointly typical with $y_1^n$ with high probability. Likewise,
the second vector quantizer can use $I(y_2;u_2)$ bits per sample to send to
its Slepian-Wolf encoder a string $u_2^n$ that is jointly typical
with $y_2^n$ with high probability.

The Slepian-Wolf encoders could view the quantized strings $u_1^n$ and
$u_2^n$ as individual symbols from a digital source to be 
compressed~\cite{Viswanathan:CEO}. They would then accumulate many
such symbols to compress. Alternatively, one
can apply the arguments behind the Slepian and Wolf theorem directly
to $u_1^n$ and $u_2^n$~\cite{Berger:MTSC,Berger:CEO,Gastpar:WZ,
Oohama:CEO:Region}. Either way, the decoder can recover 
$u_1^n$ and $u_2^n$ so long as
\begin{equation}
\begin{split}
   R_1 & \ge I(y_1;u_1|u_2) \\
   R_2 & \ge I(y_2;u_2|u_1) \\
   R_1 + R_2 & \ge I(\ve{y};\ve{u}).
\end{split}
\end{equation}
The decoder can then in principle compute the minimum 
mean-squared error (MMSE) estimate of $\ve{y}^n$ given
$\ve{u}^n$, and~(\emph{ii}) above guarantees that this
estimate will comply with the distortion constraints. By
a time-sharing argument, one can show that the rate region is convex.
This outlines the proof of the following inner bound.
\begin{prop}[Berger-Tung Inner Bound~\cite{Berger:MTSC,Tung:PHD}]
\label{innergeneralprop}
The separation-based architecture achieves the rates
\begin{equation}
\label{innergeneral}
\begin{split}
\mathcal{R}^i(d_1,d_2) = \{(R_1,R_2) : & \ \text{there exists} \
    \ve{u} \in \mathcal{U}(d_1,d_2) \ \text{such that} \\
     R_1 & \ge I(y_1;u_1|u_2) \\
     R_2 & \ge I(y_2;u_2|u_1) \\
     R_1 + R_2 & \ge I(\ve{y};\ve{u}) \}.
\end{split}
\end{equation}
In particular, the rate region contains the convex hull of this set.
\end{prop}

It is unclear \emph{a priori} how to compute this inner bound.
A natural approach is to place additional constraints on $\ve{u}$ to 
create a potentially smaller inner bound that is
amenable to explicit calculation. Let $\mathcal{U}_G(d_1,d_2)$
denote the set of $\ve{u}$ in $\mathcal{U}(d_1,d_2)$ such
that $u_j$ has zero mean and unit variance for each $j$, and
there exists a random vector $\ve{z}$ such that
\begin{enumerate}
\item[(\emph{i})] For some constants $c_1$ and $c_2$ in $[0,1)$,
\begin{align*}
u_1 & = c_1 y_1 + z_1 \\
u_2 & = c_2 y_2 + z_2,
\end{align*}
\item[(\emph{ii})] $\ve{z}$ is Gaussian and its components are independent,
\item[(\emph{iii})] $\ve{z}$ is independent of $\ve{y}$,
\item[(\emph{iv})] $E[(y_j - E[y_j|\ve{u}])^2] \le d_j$ for all $j$ in
  $\{1,2\}$.
\end{enumerate}
We will refer to a random vector $\ve{u}$ satisfying conditions
(\emph{i})-(\emph{iii}) as a \emph{distributed Gaussian test channel}
or, when there is no ambiguity, as simply a \emph{test channel}.
Note that the set of distributed
Gaussian test channels is parametrized by $c_1$ and $c_2$.

Let
\begin{equation}
\label{innergaussian}
\begin{split}
\mathcal{R}_G(d_1,d_2) = \{(R_1,R_2) : & \ \text{there exists} \
    \ve{u} \in \mathcal{U}_G(d_1,d_2) \ \text{such that} \\
     R_1 & \ge I(y_1;u_1|u_2) \\
     R_2 & \ge I(y_2;u_2|u_1) \\
     R_1 + R_2 & \ge I(\ve{y};\ve{u}) \}. 
\end{split}
\end{equation}
\begin{lemma}
\label{innerexpression}
The separation-based architecture achieves $\mathcal{R}_G(d_1,d_2)$,
which satisfies
\begin{equation}
\label{innerformula}
\mathcal{R}_G(d_1,d_2) = \mathcal{R}_1^{\star}(d_1) \cap 
   \mathcal{R}_2^{\star}(d_2) \cap \mathcal{R}_{\mathrm{sum}}^{\star}(d_1,d_2).
\end{equation}
\end{lemma}
Proposition~\ref{innergeneralprop} immediately implies that the 
separation-based architecture achieves $\mathcal{R}_G(d_1,d_2)$, 
so we only need to prove the equality in~(\ref{innerformula}).
Later, we state and prove a more general version of this equality
(Lemma~\ref{Msumsdirect} in Section~\ref{remote:section}).  
Since one can verify directly that 
$\mathcal{R}_1^{\star}(d_1)$,
$\mathcal{R}_2^{\star}(d_2)$, and
$ \mathcal{R}_{\mathrm{sum}}^{\star}(d_1,d_2)$ are convex,
it follows that $\mathcal{R}_G(d_1,d_2)$ is also convex.
Thus, time sharing does not enlarge this region.
In the remainder of the paper, whenever we consider the separation-based
architecture, we will assume that $\ve{u}$ is a distributed
Gaussian test channel.

Theorem~\ref{main:theorem} and Lemma~\ref{innerexpression}
together show that $\mathcal{R}_G(d_1,d_2)$
equals the rate region. In particular, this implies that
the separation-based scheme depicted in Fig.~\ref{natsep}
is an optimal architecture for this problem. We note that the
quadratic-Gaussian two-encoder source-coding problem is not unique
in this respect. Prior work has shown this architecture to be 
optimal for other important problems as well~\cite{Zamir:Resolution,
Berger:CEO,Gastpar:WZ,
Oohama:CEO:Region,Prabhakaran:ISIT04,Wagner:ISIT05,Wagner:MTSC}. 

In fact, the separation-based architecture achieves the rates
$$
\mathcal{R}_1^{\star}(d_1) \cap 
\mathcal{R}_2^{\star}(d_2) \cap \mathcal{R}_{\mathrm{sum}}^{\star}(d_1,d_2)
$$
even if the source is not Gaussian. Let 
$\{\check{\ve{y}}^n(i)\}_{i =1}^n$ be a 
sequence of zero-mean i.i.d.\ random vectors, not necessarily Gaussian, with
covariance matrix $\m{K}_y$. We consider the same
source-coding problem as before, but with the alternate
source $\check{\ve{y}}$ in place of $\ve{y}$. 
Let $\check{\mathcal{R}}^{i}(d_1,d_2)$ denote the
inner bound obtained from Proposition~\ref{innergeneralprop}.

\begin{prop}
\label{worstcase}
The separation-based architecture achieves the rates
\begin{equation}
\mathcal{R}_1^{\star}(d_1) \cap 
    \mathcal{R}_2^{\star}(d_2) \cap \mathcal{R}_{\mathrm{sum}}^{\star}(d_1,d_2)
\end{equation}
for the source $\check{\ve{y}}$. That is, $\check{\mathcal{R}}^{i}(d_1,d_2)$ 
contains this set.
\end{prop}
\begin{proof}
See Appendix~\ref{achievability:app}.
\end{proof}

Theorem~\ref{main:theorem} and Proposition~\ref{worstcase} together
imply that the separation-based architecture requires higher rates
to send a Gaussian source than it does to send any other source with
the same covariance.
In particular, a Gaussian source has the smallest rate region for
a given covariance matrix. This result is a
two-encoder extension of the well-known fact that a Gaussian
source has the largest rate-distortion function for a given
variance~\cite[Ex.~9.7]{Gallager:IT} (see 
Lapidoth~\cite{Lapidoth:RD} for
a stronger version). 

\section{Converse Preliminaries}
\label{CEO:section}

Oohama~\cite{Oohama:Gaussian:MTSC} determined 
the rate region when only one of the
two distortion constraints is present
\begin{align}
\mathcal{R}^\star(d_1,1) & = \mathcal{R}_1^\star(d_1) \\
\mathcal{R}^\star(1,d_2) & = \mathcal{R}_2^\star(d_2).
\end{align}
As a consequence of his result, it follows that
\begin{equation}
\label{singleDouter}
\mathcal{R}^\star(d_1,d_2) \subseteq \mathcal{R}_1^\star(d_1) \cap
    \mathcal{R}_2^\star(d_2).
\end{equation}
This outer bound is tight in a certain special
case. Let $\mathcal{D}_G$ denote the set of
matrices \m{D} such that
\begin{equation}
\label{D_G:defn}
 \m{D}^{-1} = \m{K}_y^{-1} + \m{\Lambda}
\end{equation}
for some diagonal and positive semidefinite matrix $\m{\Lambda}$.
There is a one-to-one correspondence between $\mathcal{D}_G$ and
the set of distributed Gaussian test channels. Specifically, $\m{D}$
is the covariance matrix of $\ve{y} - E[\ve{y}|\ve{u}]$, where $\ve{u}$
is a distributed Gaussian test channel with
$$
c_j^2 = \frac{\lambda_j}{1 + \lambda_j} \quad j \in \{1,2\}
$$
and $\lambda_1$ and $\lambda_2$ are defined by
$$
\m{\Lambda} = \left[ \begin{array}{cc}
\lambda_1 & 0 \\
0 & \lambda_2 
\end{array}
\right].
$$
As such, we will sometimes refer to $\m{D}$, or equivalently $\m{\Lambda}$,
as a (distributed Gaussian) test channel. Note that the mutual
information between $\ve{y}$ and $\ve{u}$ can be expressed in terms
of $\m{D}$
\begin{align}
\nonumber
I(\ve{y};\ve{u}) & = h(\ve{y}) - h(\ve{y}|\ve{u}) \\
\nonumber
     & = \frac{1}{2} \log \left((2\pi e)^2 |\m{K}_y|\right) - 
            \frac{1}{2} \log \left((2\pi e)^2 |\m{D}|\right) \\
     & = \frac{1}{2} \log \frac{|\m{K}_y|}{|\m{D}|}.
\end{align}

Let $\diag(\mathcal{D}_G)$ denote the set of distortion pairs $(d_1,d_2)$ 
such that there exists a 
\m{D} in $\mathcal{D}_G$ with top-left entry $d_1$ and bottom-right entry $d_2$.
It is straightforward to verify that $(d_1,d_2)$ is in 
$\diag(\mathcal{D}_G)$ if and only if
\begin{equation}
\label{diagcond}
\max(d_1,d_2) \le \min(1,\rho^2 \cdot \min(d_1,d_2) + 1 - \rho^2).
\end{equation}
The set $\diag(\mathcal{D}_G)$ is significant because if
$(d_1,d_2)$ is not in $\diag(\mathcal{D}_G)$, then the rate region
can be determined using existing results.
\begin{lemma}
\label{donecase}
If $(d_1,d_2)$ is not in $\diag(\mathcal{D}_G)$, then
\begin{equation}
\mathcal{R}_1^\star(d_1) \cap \mathcal{R}_2^\star(d_2) \subseteq
  \mathcal{R}_\mathrm{sum}^\star(d_1,d_2).
\end{equation}
In particular, the rate region equals
\begin{equation}
\mathcal{R}^\star(d_1,d_2) = \mathcal{R}_1^\star(d_1) \cap 
  \mathcal{R}_2^\star(d_2) \cap \mathcal{R}_{\mathrm{sum}}^\star(d_1,d_2).
\end{equation}
\end{lemma}
The proof is given in Appendix~\ref{donecaseapp}.
In light of this lemma, Lemma~\ref{innerexpression}, 
and (\ref{singleDouter}), 
it suffices to show that
when $(d_1,d_2)$ is in $\diag(\mathcal{D}_G)$, 
\begin{equation*}
\mathcal{R}^\star(d_1,d_2) \subseteq \mathcal{R}_{\mathrm{sum}}^\star(d_1,d_2).
\end{equation*}
We show this in the next section.

Our proof uses a characterization of the sum rate of the
quadratic Gaussian CEO problem. In the two-encoder version of
this problem,
encoders 1 and 2 observe $y_1$ and
$y_2$, respectively, and then communicate with a single
decoder as in the original problem. 
But now $y_1$ and $y_2$ are of the form
\begin{align*}
y_1 & = a_1 x + n_1 \\
y_2 & = a_2 x + n_2,
\end{align*}
where $x$, $n_1$, and $n_2$ are independent and Gaussian, and
the decoder estimates $x$ instead of $y_1$ and $y_2$. The
distortion measure is again the average
squared error. This problem's rate region was
determined independently by 
Oohama~\cite{Oohama:CEO:Region} and
Prabhakaran, Tse, and Ramchandran~\cite{Prabhakaran:ISIT04}\footnote{In
fact, both works solved the problem for an arbitrary number of
encoders, but this generality is not needed at this point.}.
Their result shows that the separation-based architecture
is optimal for this problem.

For our purpose, we will find it more convenient
to consider the problem
in which the decoder attempts to estimate
$\bmu^T \ve{y}$ for some given vector $\bmu$.
We call this problem the
\emph{$\bmu$-sum problem}. For some values of $\bmu$,
the $\bmu$-sum problem can be coupled to a CEO problem.
For these values of $\bmu$, it follows that the
separation-based architecture is optimal.
\begin{lemma}
\label{muCEO}
The sum rate for the $\bmu$-sum problem with
$\mu_1 \cdot \mu_2 \ge 0$ and allowable distortion $d$ equals
\begin{equation}
\label{opt}
\inf\left\{\frac{1}{2} \log \frac{|\m{K}_y|}{|\m{D}|} : 
  \m{D} \in \mathcal{D}_G \ \text{and}
 \ \bmu^T \m{D} \bmu \le d\right\}.
\end{equation}
\end{lemma}

\begin{figure}
\begin{center}
\scalebox{.90}{\input{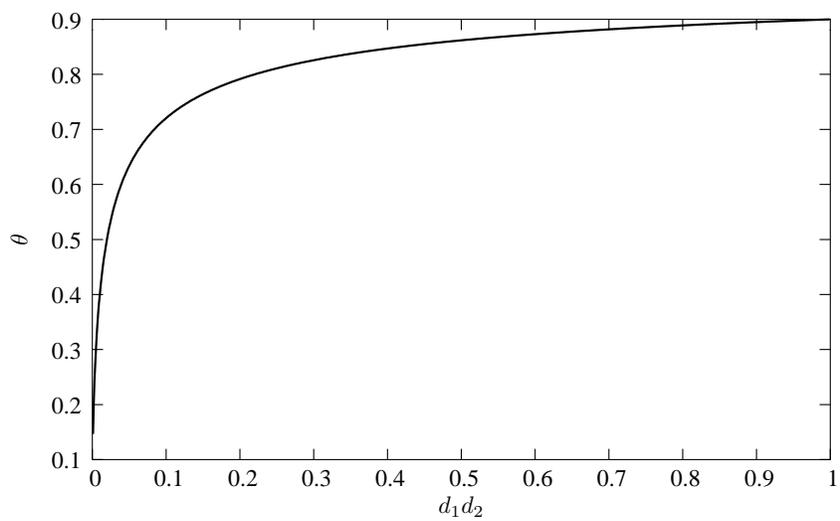}}
\end{center}
\caption{Dependence of the product $d_1 d_2$ on the error correlation
  coefficient $\theta$ for $\rho = 0.9$.}
\label{thetadepend}
\end{figure}

In Appendix~\ref{muCEOapp}, 
we prove an extended version of this lemma that includes
a description of the entire rate region.
Here we note some properties of $\mathcal{D}_G$ and 
the optimization problem~(\ref{opt}). Recall that
$\m{D}$ is in $\mathcal{D}_G$ if there exists a
diagonal and positive semidefinite matrix $\m{\Lambda}$ such that
\begin{equation}
\label{convenient}
{\m{D}}^{-1} = \m{K}_y^{-1} + \m{\Lambda}.
\end{equation}
This formula provides a convenient way of evaluating
the off-diagonal entry of $\m{D}$ in terms of its diagonal
entries and $\rho$. Let us write
\begin{equation*}
\m{D} = 
\left[
\begin{array}{cc}
d_1 & \theta \sqrt{d_1 d_2} \\
 \theta \sqrt{d_1 d_2} & d_2 
\end{array}
\right],
\end{equation*}
where $\theta \in (-1,1)$.
Equating the off-diagonal entries in~(\ref{convenient}) gives
\begin{equation}
\frac{\theta}{(1-\theta^2) \sqrt{d_1 d_2}} =
  \frac{\rho}{1 - \rho^2}.
\end{equation}
Since $\theta^2 < 1$, it follows that $\theta$ must be
positive. But this quadratic equation in $\theta$ has only
one positive root
\begin{equation}
\label{thetaroot}
\theta = \frac{\sqrt{(1-\rho^2)^2 + 4 \rho^2 d_1 d_2} - (1-\rho^2)}
  {2 \rho \sqrt{d_1 d_2}}.
\end{equation}
Thus there is no other matrix in $\mathcal{D}_G$ with top-left
entry $d_1$ and bottom-right entry $d_2$. Using~(\ref{thetaroot}),
the determinant of $\m{D}$ can be expressed in terms of $d_1$
and $d_2$
\begin{equation}
\label{detform}
|\m{D}| = \frac{2 d_1 d_2}{\beta(d_1,d_2)},
\end{equation}
where $\beta(\cdot,\cdot)$ was defined in Section~\ref{prelim:sec}.
The effect of the product $d_1 d_2$ 
on $\theta$ is shown in Fig.~\ref{thetadepend}.
As $d_1 d_2$ tends to 1, $\theta$ converges to $\rho$ and
$\m{D}$ converges to $\m{K}_y$.
On the other hand, as $d_1 d_2$ tends to zero,
$\theta$ also converges to zero, i.e., the errors become asymptotically
uncorrelated.

Next we show that every matrix in $\mathcal{D}_G$ solves
a $\bmu$-sum problem for some $\bmu$ with
$\mu_1 \cdot \mu_2 > 0$. This fact will
be used in the proof of our main result.
\begin{lemma}
\label{existsmu}
Let
$$
\m{D}^* = \left[ \begin{array}{cc}
d_1 & \theta^* \sqrt{d_1 d_2} \\
\theta^* \sqrt{d_1 d_2} & d_2 
\end{array} \right]
$$
be in $\mathcal{D}_G$, and let
\begin{equation}
\label{explicitmu}
\bmu^* = \left[ \begin{array}{c}
\sqrt{d_2} \\
\sqrt{d_1}
\end{array} \right].
\end{equation}
Then $\m{D}^*$ is sum-rate
optimal for the $\bmu^*$-sum problem, i.e.,
\begin{equation}
\frac{1}{2} \log \frac{|\m{K}_y|}{|\m{D}^*|} = 
  \inf \left\{ \frac{1}{2} \log \frac{|\m{K}_y|}{|\m{D}|} :
   \m{D} \in \mathcal{D}_G \ \text{and} \ {\bmu^*}^T \m{D} \bmu^* \le
  {\bmu^*}^T \m{D}^* \bmu^* \right\}.
\end{equation}
\end{lemma}

The proof is deferred to Appendix~\ref{existsmuapp}.
It is helpful to note that if the diagonal entries
of $\m{D}^*$ are equal, then the coordinates of $\bmu^*$
will also be equal.
This fact makes the proofs that follow somewhat simpler
when the two distortion constraints, $d_1$
and $d_2$, are equal. As such,
the reader is encouraged to keep this case in mind 
as we turn to the proof of the main result.

\section{Proof of the Main Result}
\label{proof:section}

Recall that we may restrict attention to the case in which $(d_1,d_2)$ 
is in $\diag(\mathcal{D}_G)$. Let us now fix one such distortion
pair; we will
suppress dependence on $(d_1,d_2)$ in what follows.
Let $\m{D}^*$ denote the element
of $\mathcal{D}_G$ whose top-left and bottom-right entries
are $d_1$ and $d_2$, respectively. 
\begin{defn}
For $\theta \in (-1,1)$, let
\begin{equation*}
\m{D}_\theta = \left[
\begin{array}{cc}
d_1 & \theta \sqrt{d_1 d_2} \\
\theta \sqrt{d_1 d_2} & d_2
\end{array}
\right],
\end{equation*}
and define
\begin{equation}
R_{\mathrm{coop}}(\theta) = \frac{1}{2} \log^+ \frac{|\m{K}_y|}{|\m{D}_\theta|}
  = \frac{1}{2} \log^+ \frac{1 - \rho^2}{(1-\theta^2) d_1 d_2}.
\end{equation}
Let $\bmu^*$ be the vector defined in~(\ref{explicitmu}). Then let
\begin{equation}
R_{\mathrm{sum}}(\theta) = \inf\left\{ \frac{1}{2} \log 
  \frac{|\m{K}_y|}{|\m{D}|} :
   \m{D} \in \mathcal{D}_G \ \text{and} \ {\bmu^*}^T \m{D} \bmu^*
    \le {\bmu^*}^T \m{D}_\theta \bmu^*\right\}.
\end{equation}
\end{defn}

The next lemma is central to the proof of our main result.

\begin{lemma}
\label{central}
If $(R_1,R_2,d_1,d_2)$ is strict-sense achievable, then
\begin{equation}
\label{infimum}
R_1 + R_2 \ge \inf_{\theta \in (-1,1)} \max\left(
   R_{\mathrm{coop}}(\theta),R_{\mathrm{sum}}(\theta)\right).
\end{equation}
\end{lemma}
\begin{proof}
By hypothesis there exists a code
$(f_1^{(n)},f_2^{(n)},\varphi_1^{(n)},\varphi_2^{(n)})$ 
satisfying~(\ref{feasible}). Then
\begin{align}
\nonumber
n(R_1 + R_2) & \ge H\left(f_1^{(n)}(y_1^n),f_2^{(n)}(y_2^n)\right) \\
\nonumber
    & = I\left(\mathbf{y}^n;f_1^{(n)}(y_1^n),f_2^{(n)}(y_2^n)\right) \\
\label{entropies}
    & = h(\mathbf{y}^n) - 
   h\left(\mathbf{y}^n\Big|f_1^{(n)}(y_1^n),f_2^{(n)}(y_2^n)\right),
\end{align}
where $h(\cdot)$ denotes differential entropy. But the first term
on the right-hand side satisfies
\begin{equation}
\label{firstentropy}
h(\mathbf{y}^n) = \frac{n}{2} \log \left[(2\pi e)^2 |\m{K}_y|\right]
\end{equation}
while the second term satisfies
\begin{align}
\nonumber
h\left(\mathbf{y}^n\Big|f_1^{(n)}(y_1^n),f_2^{(n)}(y_2^n)\right) & = 
  \sum_{i = 1}^n h\left(\mathbf{y}^n(i)\Big|f_1^{(n)}(y_1^n),f_2^{(n)}(y_2^n),
     \mathbf{y}^n(1:i-1)\right) \\
\nonumber
   & \le \sum_{i = 1}^n 
    h\left(\mathbf{y}^n(i) - \hat{\ve{y}}^n(i) 
            \Big|f_1^{(n)}(y_1^n),f_2^{(n)}(y_2^n)\right) \\
   & \le \sum_{i = 1}^n 
    h\left(\ve{y}^n(i) - \hat{\ve{y}}^n(i) \right),
\end{align}
since conditioning reduces entropy. Let $\hat{\m{D}}_i$ denote the
covariance matrix of $\ve{y}^n(i) - \hat{\ve{y}}^n(i)$ 
\begin{equation}
\hat{\m{D}}_i = E\left[\left(\mathbf{y}^n(i) - \hat{\ve{y}}^n(i)\right)
   \left(\mathbf{y}^n(i) - \hat{\ve{y}}^n(i)\right)^T\right],
\end{equation}
and let
\begin{equation*}
\hat{\m{D}} = \frac{1}{n} \sum_{i = 1}^n \hat{\m{D}}_i
\end{equation*}
denote the error covariance matrix of the code. We may assume that
$\varphi^{(n)}_1$ and $\varphi^{(n)}_2$ are MMSE estimators, in which case
Theorem 9.6.5 in Cover and Thomas~\cite{Cover:IT} implies that
\begin{equation*}
h\left(\mathbf{y}^n(i) - \hat{\ve{y}}^n(i)\right)
  \le \frac{1}{2} \log \left[(2 \pi e)^2 |\hat{\m{D}}_i|\right].
\end{equation*}
Applying the concavity of $\log$-$\det$~\cite[Theorem~16.8.1]{Cover:IT}, 
we have
\begin{align}
\frac{1}{n} h\left(\mathbf{y}^n\Big|f_1^{(n)}(y_1^n),f_2^{(n)}(y_2^n)\right)
   & \le \frac{1}{n} \sum_{i = 1}^n \frac{1}{2} \log
  \left[(2 \pi e)^2 |\hat{\m{D}}_i|\right] \\
   & \le \frac{1}{2} \log \left[(2 \pi e)^2 |\hat{\m{D}}|\right].
\end{align}
Combining this inequality with~(\ref{entropies}) and~(\ref{firstentropy}) 
gives
\begin{equation}
\label{coopboundbefore}
R_1 + R_2 \ge \frac{1}{2} \log^+ \frac{|\m{K}_y|}{|\hat{\m{D}}|}.
\end{equation}
Now~(\ref{entropies}) implies that
\begin{equation*}
h\left(\mathbf{y}^n\Big|f_1^{(n)}(y_1^n),f_2^{(n)}(y_2^n)\right) > - \infty.
\end{equation*}
Thus $\hat{\m{D}}$ must be nonsingular and hence positive definite.
Let us write it as
\begin{equation*}
\hat{\m{D}} = 
\left[
\begin{array}{cc}
\hat{d}_1 & \hat{\theta} \sqrt{\hat{d_1} \hat{d_2}} \\
\hat{\theta} \sqrt{\hat{d_1} \hat{d_2}} & \hat{d}_2
\end{array}
\right],
\end{equation*}
where $\hat{d}_1 \le d_1$, $\hat{d}_2 \le d_2$,
and $\hat{\theta}$ is in $(-1,1)$. Define
\begin{equation*}
\phi = \frac{\hat{\theta}{\sqrt{\hat{d}_1 \hat{d}_2}}}
   {\sqrt{d_1 d_2}},
\end{equation*}
and note that $\phi$ is in $(-1,1)$. Then $\m{D}_\phi - \hat{\m{D}}$
is diagonal
$$
\m{D}_\phi - \hat{\m{D}} = \left[ \begin{array}{cc}
d_1 - \hat{d}_1 & 0 \\
0 & d_2 - \hat{d}_2
\end{array} \right].
$$
Since $\hat{d}_1 \le d_1$ and $\hat{d}_2 \le d_2$, it follows
that $\m{D}_\phi - \hat{\m{D}}$ is positive semidefinite, i.e.,
$\hat{\m{D}} \preceq \m{D}_\phi$. In particular, 
$|\hat{\m{D}}| \le |\m{D}_\phi|$~\cite[Corollary~7.7.4]{Horn:Matrix}. This
implies that
\begin{equation}
\label{coopbound}
R_1 + R_2 \ge \frac{1}{2} \log^+ \frac{|\m{K}_y|}{|\m{D}_\phi|} = 
       R_{\mathrm{coop}}(\phi).
\end{equation}

Next observe that
\begin{equation*}
E\left[\left({\bmu^*}^T \mathbf{y}^n(i) - {\bmu^*}^T 
  \hat{\ve{y}}^n(i)\right)^2\right] = {\bmu^*}^T \hat{\m{D}}_i 
    \bmu^*.
\end{equation*}
In particular, we have
\begin{equation}
\frac{1}{n} \sum_{i = 1}^n 
  E\left[\left({\bmu^*}^T \mathbf{y}^n(i) - {\bmu^*}^T 
  \hat{\ve{y}}^n(i)\right)^2\right] 
  = {\bmu^*}^T \hat{\m{D}} \bmu^* \le {\bmu^*}^T \m{D}_\phi \bmu^*,
\end{equation}
i.e., this code achieves distortion ${\bmu^*}^T \m{D}_\phi
\bmu^*$ for the $\bmu^*$-sum problem. Lemma~\ref{muCEO} then implies that
\begin{equation*}
R_1 + R_2 \ge R_{\mathrm{sum}}(\phi).
\end{equation*}
Combining this with~(\ref{coopbound}) gives
\begin{equation*}
R_1 + R_2 \ge \max(R_{\mathrm{coop}}(\phi),R_{\mathrm{sum}}(\phi)).
\end{equation*}
The conclusion follows by taking the infimum over $\phi$ 
in $(-1,1)$.
\end{proof}

\begin{figure}
\begin{center}
\newcommand{\holdup}{\updefault}
\newcommand{\holdmd}{\mddefault}
\renewcommand{\updefault}{}
\renewcommand{\mddefault}{}
\scalebox{.65}{\input{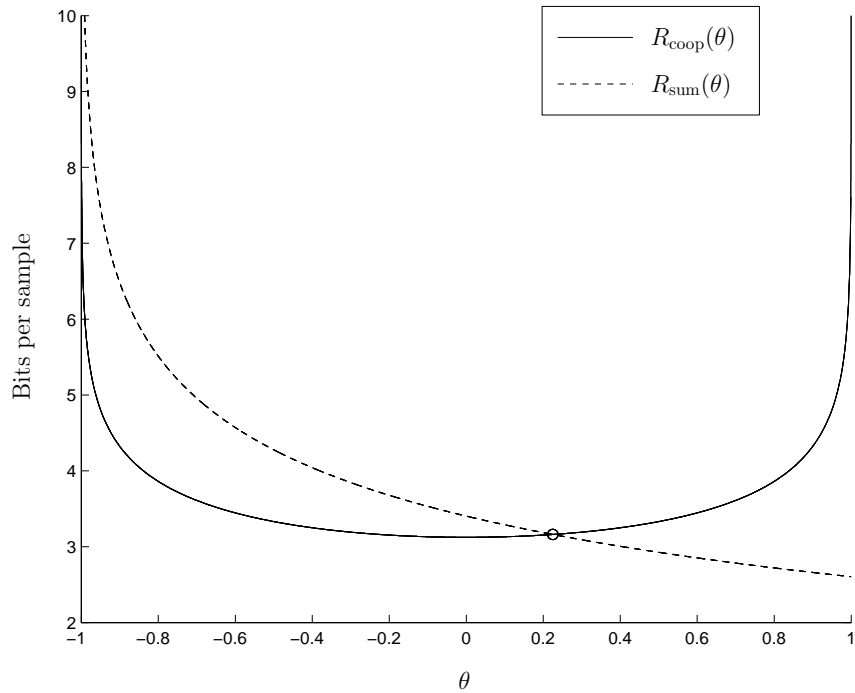}}
\renewcommand{\updefault}{\holdup}
\renewcommand{\mddefault}{\holdmd}
\end{center}
\caption{$R_{\mathrm{coop}}(\cdot)$ and $R_{\mathrm{sum}}(\cdot)$
for the case $\rho = 0.9$ and $d_1 = d_2 = 0.05$.  The plot for 
$R_{\mathrm{sum}}(\cdot)$ was generated using the convex
optimization formulation of the sum rate for the $\bmu$-sum
problem given in Appendix~\ref{muCEOapp}.
The circled point at which the two functions 
intersect is the min-max and equals the sum rate.}
\label{thetaplot}
\end{figure}

The next step is to evaluate the infimum in~(\ref{infimum}).
Examples of $R_{\mathrm{coop}}(\cdot)$ and $R_{\mathrm{sum}}(\cdot)$
are shown in Fig.~\ref{thetaplot}. We show that these two
functions always intersect at the
correlation coefficient of $\m{D}^*$, $\theta^*$,
and at this point, they equal the min-max.

\begin{lemma}
\label{calc}
\begin{align*}
\inf_{\theta \in (-1,1)} \max\left(
   R_{\mathrm{coop}}(\theta),R_{\mathrm{sum}}(\theta)\right) 
  & = R_\mathrm{coop}(\theta^*) \\
  & = R_\mathrm{sum}(\theta^*) \\
  & = \frac{1}{2} \log \frac{|\m{K}_y|}{|\m{D}^*|} \\
 &  = \frac{1}{2} \log^+ \left[ \frac{(1-\rho^2) \; \beta(d_1,d_2)}
  {2 d_1 d_2}\right].
\end{align*}
\end{lemma}
\begin{proof}
Let us write $\m{D}^*$, the matrix in $\mathcal{D}_G$ with
diagonal entries $(d_1,d_2)$, as
\begin{equation*}
\m{D}^* = 
\left[
\begin{array}{cc}
d_1 & \theta^* \sqrt{d_1 d_2} \\
\theta^* \sqrt{d_1 d_2} & d_2 \\
\end{array}
\right].
\end{equation*}
Then observe that since $\theta^* > 0$, if $\theta \ge \theta^*$,
we have
\begin{equation}
\max(R_{\mathrm{coop}}(\theta),
R_{\mathrm{sum}}(\theta)) \ge R_{\mathrm{coop}}(\theta) \ge
  R_{\mathrm{coop}}(\theta^*) = \frac{1}{2} \log \frac{|\m{K}_y|}{|\m{D}^*|}.
\end{equation}
On the other hand, if $\theta \le \theta^*$, then since
$R_{\mathrm{sum}}(\cdot)$ is nonincreasing,
\begin{equation}
\max(R_{\mathrm{coop}}(\theta),
R_{\mathrm{sum}}(\theta)) \ge R_{\mathrm{sum}}(\theta) \ge
  R_{\mathrm{sum}}(\theta^*) = \frac{1}{2} \log \frac{|\m{K}_y|}{|\m{D}^*|},
\end{equation}
where we have used the fact that 
$\m{D}^*$ solves the $\bmu^*$-sum problem. It follows that 
\begin{equation*}
\inf_{\theta \in (-1,1)} \max(R_{\mathrm{coop}}(\theta),
R_{\mathrm{sum}}(\theta)) = R_\mathrm{coop}(\theta^*) =
   R_\mathrm{sum}(\theta^*) = \frac{1}{2} \log \frac{|\m{K}_y|}{|\m{D}^*|}.
\end{equation*}
We conclude the proof by invoking the formula for the
determinant of a matrix in $\mathcal{D}_G$~(\ref{detform}).
\end{proof}

\begin{proof}[Proof of Theorem~\ref{main:theorem}]
As discussed in Section~\ref{CEO:section}, it suffices to show that
\begin{equation*}
\mathcal{R}^\star(d_1,d_2) \subseteq \mathcal{R}_{\mathrm{sum}}^\star(d_1,d_2).
\end{equation*}
Lemmas~\ref{central} and~\ref{calc} together imply that if the rate-distortion
vector $(R_1,R_2,d_1,d_2)$ is
strict-sense achievable and $(d_1,d_2)$ is in $\diag(\mathcal{D}_G)$, then
\begin{equation}
\label{bigfinish}
R_1 + R_2 \ge \frac{1}{2} \log^+ \left[ \frac{(1-\rho^2) \; \beta(d_1,d_2)}
  {2 d_1 d_2}\right].
\end{equation}
On the other hand, Lemma~\ref{donecase} implies this inequality if 
$(R_1,R_2,d_1,d_2)$ is strict-sense achievable and $(d_1,d_2)$
in not in $\diag(\mathcal{D}_G)$. It follows that~(\ref{bigfinish})
holds whenever $(R_1,R_2,d_1,d_2)$ is strict-sense achievable.
Since the right-hand side is continuous in $(d_1,d_2)$, this implies
that if the point $(R_1,R_2,d_1,d_2)$ is in $\overline{\mathcal{RD}^\star}$, 
then~(\ref{bigfinish}) again holds. This implies the desired
conclusion.
\end{proof}

\subsection{Reprise}

\begin{figure}
\begin{center}
\scalebox{0.9}{\input{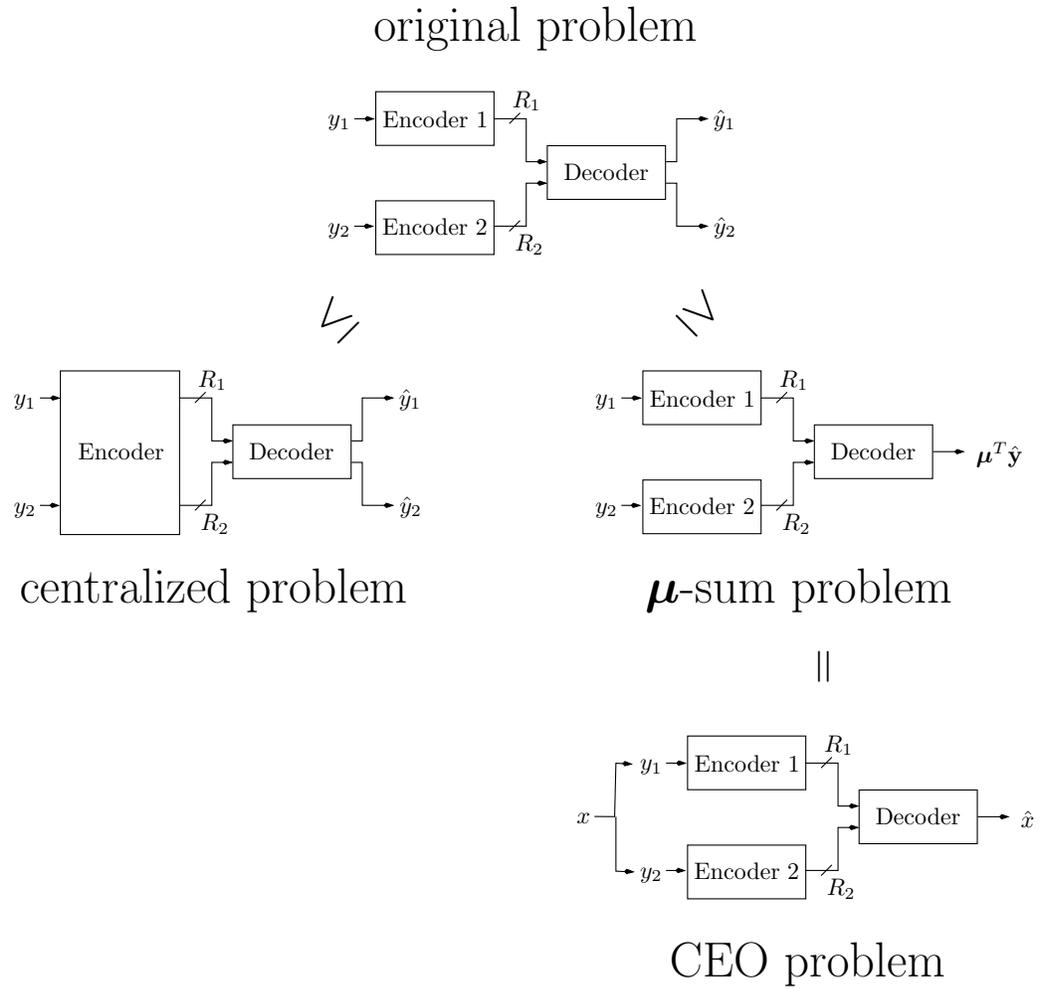}}
\end{center}
\caption{Graphical depiction of the lower bound argument.}
\label{flow:fig}
\end{figure}

The argument used in the converse proof can be summarized as
follows. Since the distortion constraints only constrain the
magnitude of the individual errors, and not their correlation,
we view the determination of the sum-rate as an implicit
minimization over all possible error covariance matrices, 
subject to upper bounds on the diagonal elements. We then
lower bound the sum rate for each possible error covariance matrix
using two approaches. First, we consider the rate needed by
a centralized encoder to achieve the given error covariance matrix.
Second, we use the existing characterization
of the rate region for the CEO problem to solve the $\bmu$-sum
problem for some $\bmu$ vectors.
This solution is then used to lower bound the sum rate of
the problem under study for a given error covariance matrix.
The first
bound is most effective when the correlation between the errors
is large. The second bound is most effective when the correlation
is small. We therefore form a composite bound by taking the 
maximum of these two lower bounds. The argument is illustrated
in Figs.~\ref{thetaplot} and~\ref{flow:fig}. Note that both of the lower
bounds are needed.

\section{The $\m{M}$-sums Problem}
\label{remote:section}

Consider next a generalization of the classical problem
in which the decoder attempts to estimate $\bmu_j^T \ve{y}$
for a given set of vectors $\bmu_1,\ldots,\bmu_J$. We
may assume without loss of generality that these vectors are
distinct and have unit norm.

Define the matrix
\begin{equation*}
\m{M} = [ \bmu_1 \ \bmu_2 \cdots \bmu_J]
\end{equation*}
consisting of the column vectors $\bmu_1, \ldots, \bmu_J$ side-by-side.
The problem is then to reproduce the vector $\m{M}^T \ve{y}$ 
subject to separate constraints on the average squared error of 
each component. We call this the \emph{$\m{M}$-sums problem.}
Note that the classical quadratic Gaussian two-encoder source coding
problem can be viewed as an instance of the $\m{M}$-sums problem
with $\m{M}$ equal to the identity matrix. We will
show that the techniques used to solve that
problem can be used to solve the general $\m{M}$-sums problem 
if the vectors $\bmu_1, \ldots \bmu_J$ satisfy a certain
condition. Specifically, we will require that the product of
the two coordinates of each vector is nonnegative
\begin{equation}
\label{Msumscondition}
\bmu_{j1} \cdot \bmu_{j2} \ge 0 \quad \forall j \in \{1,\ldots,J\}.
\end{equation}
This condition is satisfied if and only for each $j$ either both coordinates
of $\bmu_j$ are nonnegative or both are nonpositive. From a source-coding
perspective, these two cases are essentially equivalent, so for
simplicity
we will assume that the components of $\bmu_j$ are nonnegative
for each $j$.

The condition in~(\ref{Msumscondition}) depends on our standing assumption
that $0 < \rho < 1$. If $\rho$ is negative, then the 
condition in~(\ref{Msumscondition}) becomes
\begin{equation}
\label{Msumscondition2}
\bmu_{j1} \cdot \bmu_{j2} \le 0 \quad \forall j \in \{1,\ldots,J\}.
\end{equation}
Note that either way, the condition includes the case when $\m{M}$
is the identity matrix, i.e., the classical version of the problem.

\subsection{Main Result}

In this section, we use $\mathcal{R}_j^\star(d_j)$ to denote
the rate region of the $\bmu_j$-sum problem with distortion
constraint $d_j$. Let $R_\mathrm{sum}^\star(d_1,\ldots,d_J)$ denote the
minimum sum rate for the $\m{M}$-sums problem achieved by 
the separation-based scheme
\begin{multline}
R_\mathrm{sum}^\star(d_1,\ldots, d_J) \\
  = \inf \left\{ \frac{1}{2} \log
   \frac{|\m{K}_y|}{|\m{D}|} : \m{D} \in \mathcal{D}_G \ \text{and} \
      \bmu_j^T \m{D} \bmu_j \le d_j \ \forall j \in \{ 1,\ldots,J\}\right\}.
\end{multline}
Then let $\mathcal{R}^\star_\mathrm{sum}(d_1,\ldots,d_J)$ denote the
set of rate pairs whose sum is at least 
 $R_\mathrm{sum}^\star(d_1,\ldots,d_J)$ 
\begin{equation}
\mathcal{R}_\mathrm{sum}^\star(d_1,\ldots,d_J) = \{(R_1,R_2) :
  R_1 + R_2 \ge  R_\mathrm{sum}^\star(d_1,\ldots,d_J) \}.
\end{equation}
In terms of these sets, the separation-based architecture
achieves the following inner bound.
\begin{lemma}
\label{Msumsdirect}
For the $\m{M}$-sums problem, the separation-based architecture
achieves the rates
\begin{equation*}
\mathcal{R}^\star_\mathrm{sum}(d_1,\ldots,d_J) \cap
  \bigcap_{j = 1}^J \mathcal{R}_j^\star(d_j).
\end{equation*}
\end{lemma}
The proof is elementary but somewhat involved and is
given in Appendix~\ref{Msumsdirectapp}. The main result of this
section is the following theorem that shows that this
inner bound equals the rate region.
\begin{theorem}
\label{Msums:theorem}
The rate region of the $\m{M}$-sums problem equals
\begin{equation}
\label{Msumsregion}
\mathcal{R}_\mathrm{sum}^\star(d_1,\ldots,d_J) \cap
  \bigcap_{j = 1}^J \mathcal{R}_j^\star(d_j).
\end{equation}
\end{theorem}
The proof parallels that of Theorem~\ref{main:theorem}. In particular,
we use functions similar to $R_\mathrm{coop}(\cdot)$ and
$R_\mathrm{sum}(\cdot)$. 
The details are given in Appendix~\ref{Msumsapp}.

By mimicking the proof of Proposition~\ref{worstcase}, one can show that
the separation-based architecture achieves the rates 
in~(\ref{Msumsregion}) even if the source is not Gaussian.
Theorem~\ref{Msums:theorem} then implies that, as with the
classical version of the problem, the separation-based
inner bound and the rate region are both smallest for a Gaussian source.

\subsection{The Remote-source Problem}

As an application of Theorem~\ref{Msums:theorem}, consider the remote-source
version of the original
problem. Here the encoders' observations are viewed as an
underlying source, $\tilde{\ve{y}}$, plus additive noise
\begin{align*}
y_1 & = \tilde{y}_1 + n_1 \\
y_2 & = \tilde{y}_2 + n_2
\end{align*}
where $\tilde{\ve{y}}$, $n_1$, and $n_2$ are independent 
and Gaussian. We assume these random variables have zero
mean and the following second moments
\begin{align*}
E[\tilde{y}_1^2] & = \sigma^2_1 \le 1 \\
E[\tilde{y}_2^2] & = \sigma^2_2 \le 1 \\
E[\tilde{y}_1 \tilde{y}_2] & = \rho \\
E[n_1^2] & = 1 - \sigma_1^2 \\
E[n_2^2] & = 1 - \sigma_2^2
\end{align*}
so that $\ve{y}$ has covariance matrix $\m{K}_y$. The aim is
to reproduce $\tilde{y}_1$ and $\tilde{y}_2$ subject to distortion
constraints $d_1$ and $d_2$, respectively.
A partial characterization of the
rate region for this problem was obtained by 
Oohama~\cite{Oohama:Remote:Allerton}.
By coupling this problem to an $\m{M}$-sums problem, we can
determine the rate region completely.
\begin{cor}
\label{remotecor}
The rate region for the remote-source problem with
distortion constraints $d_1$ and $d_2$ 
equals the rate region for the
$\m{M}$-sums problem with distortion constraints
$d_1 - \gamma_1$ and $d_2 - \gamma_2$, where
\begin{equation*}
\m{M} = \frac{1}{1 - \rho^2} \left[
\begin{array}{cc}
\sigma_1^2 - \rho^2 & \rho(1 - \sigma_2^2) \\
\rho(1 - \sigma_1^2) & \sigma_2^2 - \rho^2 
\end{array}
\right] = [\bmu_1 \ \ \bmu_2]
\end{equation*}
and
\begin{align*}
\gamma_1 & = \frac{(\sigma_1^2 - \rho^2) (1 - \sigma^2_1)}
   {1 - \rho^2} \\
\gamma_2 & = \frac{(\sigma_2^2 - \rho^2) (1 - \sigma^2_2)}
   {1 - \rho^2}.
\end{align*}
\end{cor}
\begin{proof}
Standard calculations show that $E[\tilde{y}_1|\ve{y}] = \bmu_1^T \ve{y}$,
and in particular, $\tilde{y}_1$ can be written
\begin{equation*}
\tilde{y}_1 = \bmu_1^T \ve{y} + \tilde{n}_1,
\end{equation*}
where $\tilde{n}_1$ is Gaussian, independent of $\ve{y}$, and has
mean zero and
variance $\gamma_1$. Then for any random variable $\overline{\ve{y}}$ 
such that
$\tilde{\ve{y}} \markov \ve{y} \markov \overline{\ve{y}}$, we have
\begin{align}
\nonumber
E[(\tilde{y}_1 - \overline{y}_1)^2] & = 
\nonumber
E[(\tilde{y}_1 - \bmu_1^T \ve{y} + \bmu_1^T \ve{y} - \overline{y}_1)^2] \\
\nonumber
& = E[(\tilde{n}_1 + \bmu_1^T \ve{y} - \overline{y}_1)^2]  \\
\label{coupleddistortion}
& = \gamma_1 + E[(\bmu_1^T \ve{y} - \overline{y}_1)^2].
\end{align}
Now consider a pair of encoders, $f_1^{(n)}$ and $f_2^{(n)}$,
and a decoder $(\varphi_1^{(n)},\varphi_2^{(n)})$ with
$$
\overline{y}^n_i = \varphi_i^{(n)}\left(f_1^{(n)}(y_1^n),f_2^{(n)}(y_2^n)
   \right) \quad
  i = 1,2.
$$
Since the source is i.i.d., for any time $i$, we have
$$
\tilde{\ve{y}}^n(i) \markov \ve{y}^n(i) \markov 
   \overline{\ve{y}}^n(i).
$$
Thus by~(\ref{coupleddistortion}), it follows that
\begin{equation}
E\left[(\tilde{y}_1^n(i) - \overline{y}_1^n(i))^2\right]
  = \gamma_1 + E[(\bmu_1^T\ve{y}^n(i) - \overline{y}_1^n(i))^2].
\end{equation}
By averaging both sides of this equation over time, we see that any code
that achieves distortion $d_1$ for $\tilde{y}_1$ must achieve distortion
$d_1 - \gamma_1$ for $\bmu_1^T \ve{y}$ and vice versa. Likewise, any code
that achieves distortion $d_2$ for $\tilde{y}_2$ must achieve distortion
$d_2 - \gamma_2$ for $\bmu_2^T \ve{y}$ and vice versa. The conclusion follows.
\end{proof}

Observe that this proof does not require the assumption that
$0 < \rho < 1$, only that $\rho^2 < 1$. If $0 < \rho < 1$, then
$\bmu_1$ and $\bmu_2$ satisfy the condition
in~(\ref{Msumscondition}). On the other hand, if $-1 < \rho < 0$,
then $\bmu_1$ and $\bmu_2$ satisfy the condition in~(\ref{Msumscondition2}).
Since the cases $\rho = 0$, $\rho = -1$, and $\rho = 1$ can be 
solved using existing techniques, the rate region for the remote
source problem is solved for any value of $\rho$.

\section{Many Sources}
\label{many:section}

\begin{figure}
\begin{center}
\scalebox{1.1}{\input{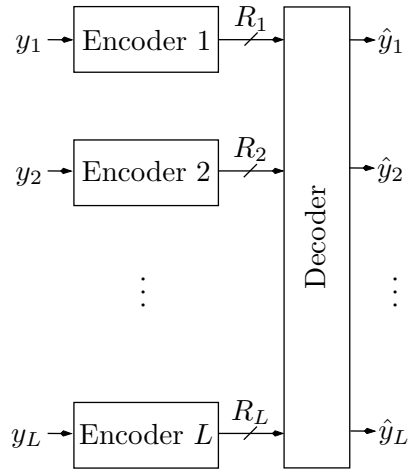}}
\end{center}
\caption{Setup for the many sources problem.}
\label{L:fig}
\end{figure}

Our technique can be used to determine the sum rate 
for more than two sources
if a certain symmetry condition holds. Suppose 
now that there are $L$ jointly Gaussian sources,
$y_1, \ldots, y_L$, with covariance matrix
\begin{equation*}
\m{K}_y = \left[
\begin{array}{cccc}
1 & \rho & \cdots & \rho \\
\rho & 1 & \ldots & \rho \\
\vdots &  \vdots & \ddots & \vdots \\
\rho & \rho & \cdots & 1
\end{array}
\right]
\end{equation*}
for some $0 < \rho < 1$. That is, the source 
components are Gaussian, exchangeable, and positively correlated.
We assume that the sources are
separately encoded, as shown in Fig.~\ref{L:fig}, and that
$L$ distortion constraints are imposed on the individual
reproductions
\begin{equation*}
d_\ell \ge \frac{1}{n} \sum_{i = 1}^n E[(y_\ell^n(i) - \hat{y}_\ell^n(i))^2]
  \quad \text{for all $\ell$ in $\{1,\ldots,L\}$}.
\end{equation*}
The separation-based scheme yields an inner bound on the rate region,
and in particular, an upper bound on the sum rate.  As in the 
case of two sources, let $\mathcal{D}_G$ denote the set of
matrices $\m{D}$ such that
$$
\m{D}^{-1} = \m{K}_y^{-1} + \m{\Lambda}
$$
for some diagonal and positive semidefinite matrix $\m{\Lambda}$.
The sum rate achieved by the separation-based scheme is then
\begin{equation}
\label{manysourcesum}
\inf\left\{ \frac{1}{2} \log \frac{|\m{K}_y|}{|\m{D}|} : \m{D} \in 
   \mathcal{D}_G \ \text{and} \ \ve{e}_\ell^T 
  \m{D} \ve{e}_\ell \le d_\ell \ \forall \
    \ell \in \{1,\ldots,L\}\right\},
\end{equation}
where $\ve{e}_\ell$ denotes the vector with one in position
$\ell$ and zero elsewhere. By following the proof of 
Theorem~\ref{main:theorem},
one can show that this sum rate is optimal if the distortion constraints
$d_1,\ldots,d_L$ are equal.

\begin{theorem}
\label{manysource:theorem}
If $d_1 = d_2 = \cdots = d_L = d$, then the
separation-based architecture is sum-rate optimal. In particular,
the sum rate is given by~(\ref{manysourcesum}). Furthermore, 
in this case the 
infimum in~(\ref{manysourcesum}) is achieved by a
$\m{D}$ in $\mathcal{D}_G$ of the form
$$
\m{D}^{-1} = \m{K}_y^{-1} + \lambda \m{I},
$$
for some $\lambda \ge 0$.
\end{theorem}

The proof is given in Appendix~\ref{manyapp}. As
with the $\m{M}$-sums problem, it is possible to
mimic Proposition~\ref{worstcase} and show that the 
separation-based architecture achieves the
sum rate in (\ref{manysourcesum}) even if the source is 
not Gaussian. It follows that the Gaussian
source has the largest sum rate among all
exchangeable and positively correlated sources
when all of the distortion constraints are equal.

\section{Concluding Remarks}
\label{remarks:section}

We determined the rate region of the quadratic Gaussian
two-encoder source-coding problem. This result implies that
a simple architecture that separates the analog and digital
aspects of the compression is optimal, and
that this architecture requires higher rates to send a Gaussian
source than it does to send any other source with the same
covariance. We also described how our proof technique can
be extended to determine the sum rate of some generalizations
of this problem. We now comment on two aspects of our results.

\subsection{An Extremal Result}

One consequence of our main result is that there is no loss of
optimality in using Gaussian auxiliary random variables 
in the separation-based inner bound.
More precisely, the regions $\mathcal{R}^i(d_1,d_2)$ and
$\mathcal{R}_G(d_1,d_2)$ defined in~(\ref{innergeneral}) and
(\ref{innergaussian}) are equal. In
particular, these two regions have the same sum rate.
Thus to the optimization problem
\begin{equation}
\begin{split}
\text{minimize} & \quad I(\ve{y};\ve{u}) \\
\text{subject to} & \quad u_1 \markov y_1 \markov y_2 \markov u_2 \\
  &  \quad  E[(y_j - E[y_j|\ve{u}])^2] \le d_j \quad j \in \{1,2\}
\end{split}
\end{equation}
we can add the constraint
\begin{equation*}
(\ve{y},\ve{u}) \ \text{is jointly Gaussian}
\end{equation*}
without changing the optimal value.
The same is true, of course, of the optimization problem
\begin{equation}
\begin{split}
\text{maximize} & \quad h(\ve{y}|\ve{u}) \\
\text{subject to} & \quad u_1 \markov y_1 \markov y_2 \markov u_2 \\
  & \quad  E[(y_j - E[y_j|\ve{u}])^2] 
     \le d_j \quad j \in \{1,2\}.
\end{split}
\end{equation}
This is akin to the well-known fact that the Gaussian
distribution maximizes entropy for a given covariance.
But this result is more subtle in that the conditional
covariance of $\ve{y}$ given $\ve{u}$ is not fixed, and
by using non-Gaussian $\ve{u}$, one can potentially realize
conditional covariances that are unattainable with Gaussian
distributions. Evidently the entropy-maximizing property of
the Gaussian distribution more than compensates for its
smaller set of achievable conditional covariances.

Using Theorem~\ref{Msums:theorem}, it is possible to generalize this
result to allow distortion constraints on linear combinations
of the source variables $y_1$ and $y_2$. 
It is also possible to prove a multi-letter
version of this result by first proving a multi-letter version
of the inner bound, in which several source symbols are treated
as a single ``supersymbol.'' Whether one can
prove any of these extremal results
without reference to the source-coding
setup that is the subject of this paper is an interesting open
question.

\subsection{Source Augmentation}

The most noteworthy aspect of our proof
is the random variable
$x$ that we add to the source $\ve{y}$ in Appendix~\ref{muCEOapp}
to solve the $\bmu$-sum problem.
Unlike other more typical auxiliary random variables, $x$
does not represent a component of the code. Rather, it is
used to aid the analysis by inducing conditional independence among
the observations, which allows us to couple our problem to
a CEO problem. Of course, there are many random variables that
will induce conditional independence. The role of Lemma~\ref{existsmu}
is to identify the best one.

This technique of augmenting
the source to induce conditional independence has proven
useful in other contexts as well. Ozarow~\cite{Ozarow:MD} used
it to prove the converse for the Gaussian two-descriptions problem.
Wang and Viswanath~\cite{Wang:MD} used it to determine the sum rate
for the Gaussian vector multiple-descriptions problem with
individual and central decoders. Wagner and 
Anantharam~\cite{Wagner:ISIT05,Wagner:MTSC} used it to prove an outer 
bound for the discrete multiterminal source-coding problem.

Recently, we have generalized the CEO result to sources
whose correlation satisfies a certain tree condition~\cite{Tavildar:ITW06}
(see also~\cite{Oohama:ISIT:Tree}).
This suggests an approach for generalizing the results in
this paper. Specifically, one could potentially augment the source
to couple a given distributed source coding problem to this tree problem
instead of the more restrictive CEO problem. Determining whether 
this revised approach yields stronger results is a worthwhile
question for future research.

\section*{Acknowledgment}

We wish to thank Venkat Anantharam and Jun Chen for helpful discussions.
We would also like to thank the anonymous reviewers for carefully
checking the manuscript and suggesting many improvements.

\appendix

\section{Proof of Proposition~\ref{worstcase}}
\label{achievability:app}

Let $(R_1,R_2)$ be a rate pair in 
$$
\mathcal{R}_1^\star(d_1) \cap
   \mathcal{R}_2^\star(d_2) \cap \mathcal{R}_{\mathrm{sum}}^\star(d_1,d_2).
$$
By Lemma~\ref{innerexpression}, there exists a $\ve{u}$ in
$\mathcal{U}_G(d_1,d_2)$ such that
\begin{equation}
\begin{split}
R_1 & \ge I(y_1;u_1|u_2) \\
R_2 & \ge I(y_2;u_2|u_1) \\
R_1 + R_2 & \ge I(\ve{y};\ve{u}).
\end{split}
\end{equation}
Now $\ve{u}$ can be expressed as
\begin{align*}
u_1 & = c_1 y_1 + z_1 \\
u_2 & = c_2 y_2 + z_2 
\end{align*}
for some coefficients $c_1$ and $c_2$ in $[0,1)$, 
where $z_1$, $z_2$,
and $\ve{y}$ are independent and $\ve{z}$ is Gaussian. Now
construct auxiliary random variables $\check{u}_1$ and
$\check{u}_2$ for the true source via
\begin{align*}
\check{u}_1 & = c_1 \check{y}_1 + z_1 \\
\check{u}_2 & = c_2 \check{y}_2 + z_2 
\end{align*}
with $\ve{z}$ independent of $\check{\ve{y}}$. Note
that the Markov condition
\begin{equation*}
\check{u}_1 \markov \check{y}_1 \markov \check{y}_2 \markov
   \check{u}_2
\end{equation*}
is satisfied and that $(\check{\ve{y}},\check{\ve{u}})$ and
$(\ve{y},\ve{u})$ have the same second-order statistics. Thus the
error in the linear minimum mean-squared error estimate of
$\check{y}_j$ given $\check{\ve{u}}$, 
$\mathrm{LMMSE}(\check{y}_j|\check{\ve{u}})$, equals the error
in the linear minimum mean-squared error estimate of $y_j$
given $\ve{u}$,
\begin{equation}
   \mathrm{LMMSE}(\check{y}_j|\check{\ve{u}}) = \mathrm{LMMSE}(y_j|\ve{u}) 
   \quad j \in \{1,2\}.
\end{equation}
But conditional expectation minimizes mean square error,
\begin{equation}
E[(\check{y}_j - E[\check{y}_j|\check{\ve{u}}])^2] \le 
   \mathrm{LMMSE}(\check{y}_j|\check{\ve{u}}) \quad j \in \{1,2\},
\end{equation}
and for jointly Gaussian random variables, the 
linear minimum mean-squared error estimate is also the
conditional expectation~\cite[Theorems 9.1-1 and 9.1-2]{Stark:Woods:3}. 
Since $\ve{u}$ is in $\mathcal{U}_G(d_1,d_2)$,
this implies
\begin{equation*}
E[(\check{y}_j - E[\check{y}_j|\check{\ve{u}}])^2] \le d_j \quad 
       j \in \{1,2\}.
\end{equation*}
It follows that $\check{\ve{u}}$ is in $\mathcal{U}(d_1,d_2)$.
Next, we show that $(R_1,R_2)$ satisfies
\begin{equation}
\begin{split}
R_1 & \ge I(\check{y}_1;\check{u}_1|\check{u}_2) \\
R_2 & \ge I(\check{y}_2;\check{u}_2|\check{u}_1) \\
R_1 + R_2 & \ge I(\check{\ve{y}};\check{\ve{u}}).
\end{split}
\end{equation}
To prove this, it suffices to show that
\begin{align}
\nonumber
I(y_1;u_1|u_2) & \ge I(\check{y}_1;\check{u}_1|\check{u}_2) \\
\nonumber
I(y_2;u_2|u_1) & \ge I(\check{y}_2;\check{u}_2|\check{u}_1) \\
\label{sumdom}
I(\ve{y};\ve{u}) & \ge I(\check{\ve{y}};\check{\ve{u}}).
\end{align}
By symmetry, it suffices to prove the last two inequalities.  Let
\begin{equation*}
\alpha_2 = \frac{E[u_1 u_2]}{E[u_1^2]} = E[u_1 u_2].
\end{equation*}
Then we have
\begin{align*}
I(\check{y}_2;\check{u}_2|\check{u}_1) & \stackrel{(a)}{=}
     h(\check{u}_2|\check{u}_1) - h(\check{u}_2|\check{y}_2) \\
     & \stackrel{(b)}{=} h(\check{u}_2 - \alpha_2 \check{u}_1|\check{u}_1) 
           - h(z_2) \\
     & \stackrel{(c)}{\le} h(\check{u}_2 - \alpha_2 \check{u}_1) 
           - h(z_2) \\
     & \stackrel{(d)}{\le} h(u_2 - \alpha_2 u_1) 
           - h(z_2) \\
     & \stackrel{(e)}{=} I(y_2;u_2|u_1), 
\end{align*}
where
\begin{enumerate}
\item[(\emph{a})] follows from the Markov condition $\check{u}_2
  \markov \check{y}_2 \markov \check{u}_1$,
\item[(\emph{b})] follows because differential entropy is invariant
   to shifts,
\item[(\emph{c})] follows because conditioning reduces differential entropy,
\item[(\emph{d})] follows from the fact that $\ve{u}$ has the same
covariance as $\check{\ve{u}}$, and the
Gaussian distribution maximizes differential entropy for a given
variance, and
\item[(\textit{e})] follows because steps (\textit{c}) and 
   (\textit{d}) are tight if $(\check{\ve{y}},\check{\ve{u}})$ is Gaussian.
\end{enumerate}
We can prove~(\ref{sumdom}) via similar reasoning
\begin{equation}
\begin{split}
I(\check{\ve{y}};\check{\ve{u}}) & = h(\check{\ve{u}}) -
    h(\check{\ve{u}}|\check{\ve{y}}) \\
      & = h(\check{\ve{u}}) - h(\ve{z}) \\
      & \le h(\ve{u}) - h(\ve{z}) \\
       & = I(\ve{y};\ve{u}).
\end{split}
\end{equation}
It follows that the rate pair $(R_1,R_2)$ belongs to 
$\check{\mathcal{R}}^i(d_1,d_2)$.

\section{Converse for a Special Case}
\label{donecaseapp}

\begin{proof}[Proof of Lemma~\ref{donecase}]
The conclusion is easily verified if $\min(d_1,d_2) \ge 1$, so
assume instead that $\min(d_1,d_2) < 1$. Without loss of
generality, let us assume that $d_1 = \min(d_1,d_2)$. 
Then by~(\ref{diagcond}), we must have
\begin{equation*}
\rho^2 d_1 + 1 - \rho^2 < d_2.
\end{equation*}
If this holds, then
\begin{equation*}
\mathcal{R}_{\mathrm{sum}}^\star(d_1, \rho^2 d_1 + 1 - \rho^2) \subseteq
  \mathcal{R}_{\mathrm{sum}}^\star(d_1,d_2).
\end{equation*}
But one can verify directly that 
\begin{equation}
\mathcal{R}_{\mathrm{sum}}^\star(d_1, \rho^2 d_1 + 1 - \rho^2)
  = \left\{(R_1,R_2) : R_1 + R_2 \ge \frac{1}{2} \log \frac{1}{d_1} \right\}.
\end{equation}
Now via calculus one can show that 
if $(R_1,R_2)$ is in $\mathcal{R}_1^\star(d_1)$ then $R_1$ and
$R_2$ must satisfy
\begin{equation*}
R_1 + R_2 \ge \frac{1}{2} \log \frac{1}{d_1}.
\end{equation*}
It follows that
\begin{equation*}
\mathcal{R}_1^\star(d_1) \subseteq \mathcal{R}_{\mathrm{sum}}^\star(d_1,d_2).
\end{equation*}
In particular, we have
\begin{equation}
\mathcal{R}_1^\star(d_1) \cap \mathcal{R}_2^\star(d_2) =
   \mathcal{R}_1^\star(d_1) \cap \mathcal{R}_2^\star(d_2) \cap
   \mathcal{R}_{\mathrm{sum}}^\star(d_1,d_2).
\end{equation}
The result then follows from~(\ref{singleDouter}) and 
Lemma~\ref{innerexpression}.
\end{proof}

\section{The $\bmu$-sum Problem}
\label{muCEOapp}

In this appendix, we determine the rate region for the
$\bmu$-sum problem if $\bmu$ satisfies $\mu_1 \cdot \mu_2 \ge 0$.
If $\mu_1 \cdot \mu_2 = 0$, then the rate region has already
been determined by Oohama~\cite{Oohama:Gaussian:MTSC}, so we shall assume that
$\mu_1 \cdot \mu_2 > 0$.

We begin by noting that
if $\bmu$ and the allowable distortion are both scaled
by the same factor, then the rate region remains unchanged. We may therefore
assume that $\bmu$ is normalized. In
particular, we may assume that
\begin{equation}
\label{normalized}
\mu_1 \cdot \mu_2 = \frac{\gamma^2}{\rho} \cdot \SNR_1 \cdot \SNR_2
\end{equation}
where
\begin{align}
\label{SNR1}
\SNR_1 & = \frac{\rho}{1 - \rho^2} \left(\frac{\mu_1}{\mu_2} + \rho\right) \\
\label{SNR2}
\SNR_2 & = \frac{\rho}{1 - \rho^2} \left(\frac{\mu_2}{\mu_1} + \rho\right) \\
\intertext{and}
\label{gammadef}
\gamma^{-1} & = 1 + \SNR_1 + \SNR_2.
\end{align}
This normalization is convenient because, as we shall see,
it admits a particularly
simple coupling to a CEO problem.

\begin{lemma}
\label{muCEOextended}
Suppose the vector $\bmu$ satisfies $\mu_1 \cdot \mu_2 > 0$ and
the normalization~(\ref{normalized}). Then
the rate region for the $\bmu$-sum problem with allowable distortion
$d$ equals
\begin{equation}
\label{muregion}
\begin{split}
\Bigg\{(R_1,R_2) : & \ \text{there exist} \ r_1 \ge 0, r_2 \ge 0 \ 
     \text{such that} \\
     R_1 & \ge \frac{1}{2} \log^+ \left[\frac{1}{d + \gamma}
                (1 + \SNR_2(1 - 2^{-2r_2}))^{-1}\right] + r_1 \\
     R_2 & \ge \frac{1}{2} \log^+ \left[\frac{1}{d + \gamma}
                (1 + \SNR_1(1 - 2^{-2r_1}))^{-1}\right] + r_2 \\
     R_1 + R_2 & \ge \frac{1}{2} \log^+\left[\frac{1}{d + \gamma}\right]
                  + r_1 + r_2 \\
    \frac{1}{d + \gamma} & \le 1 + \sum_{j = 1}^2 \SNR_j(1 - 2^{-2r_j})
              \Bigg\}.
\end{split}
\end{equation}
In particular, the sum rate equals
\begin{equation}
\label{convexprob}
\begin{gathered}
\inf\Bigg\{\frac{1}{2}\log^+ \left[
  \frac{1}{d + \gamma}\right] + r_1 + r_2 : 
     r_1 \ge 0, \ r_2 \ge 0, \ \text{and} \\
   1 + \sum_{j = 1}^2 
   \SNR_j \left(1-2^{-2r_j}\right) \ge 
   \frac{1}{d + \gamma}\Bigg\},
\end{gathered}
\end{equation}
or, equivalently,
\begin{equation}
\label{optagain}
\inf\left\{\frac{1}{2} \log \frac{|\m{K}_y|}{|\m{D}|} : 
  \m{D} \in \mathcal{D}_G \ \text{and}
 \ \bmu^T \m{D} \bmu \le d\right\}.
\end{equation}
Furthermore, the infimum in~(\ref{optagain}) is achieved by
a unique feasible $\m{D}$.
\end{lemma}

\begin{proof}
Let 
\begin{equation}
\label{adef}
a_j = \left(\frac{\SNR_j}{1 + \SNR_j}\right)^{1/2} \quad j \in \{1,2\}.
\end{equation}
Clearly $a_1 < 1$ and $a_2 < 1$. Using (\ref{SNR1}) and
(\ref{SNR2}), one can verify that $a_1 a_2 = \rho$. It follows
that $a_1$ and $a_2$ are each contained in $(\rho,1)$.
Let $x$, $n_1$, and $n_2$ be independent zero-mean Gaussian random
variables with variances
\begin{align*}
E[x^2] & = 1 \\
E[n_j^2] & = 1 - a_j^2 \ \ \ j \in \{1,2\}.
\end{align*}
Since $(a_1 x + n_1, a_2 x + n_2)$ has covariance matrix $\m{K}_y$, we
can couple these variables to $\mathbf{y}$ to create a CEO problem
\begin{equation}
\begin{split}
\label{firstcoupling}
y_1 & = a_1 x + n_1 \\
y_2 & = a_2 x + n_2. 
\end{split}
\end{equation}
The $\SNR$ notation is justified by the fact that
\begin{equation*}
\SNR_j = \frac{a_j^2}{1 - a_j^2} = \frac{\Var(a_j x)}{\Var(n_j)}
   \quad j \in \{1,2\}.
\end{equation*}
Now starting with~(\ref{normalized}), we have
$$
\mu_1 \cdot \mu_2 = \sqrt{\mu_1 \mu_2} \cdot \frac{\gamma}{\sqrt{\rho}} \cdot
    \sqrt{\SNR_1 \SNR_2}.
$$
Substituting for $\SNR_1$ and $\SNR_2$ and rearranging gives
\begin{equation}
\label{muSNR}
\mu_1 = \gamma \cdot \sqrt{\frac{\rho \mu_1 + \rho^2 \mu_2}{\mu_2 +
    \rho \mu_1}} \cdot \frac{\mu_2 + \rho \mu_1}{(1-\rho^2){\mu_2}}.
\end{equation}
But observe that
\begin{equation*}
\frac{\mu_2 + \rho \mu_1}{(1-\rho^2){\mu_2}} = 
  1 + \SNR_1 = \frac{1}{1 - a_1^2}
\end{equation*}
and
\begin{equation*}
\sqrt{\frac{\rho \mu_1 + \rho^2 \mu_2}{\mu_2 + \rho \mu_1}}
  = a_1.
\end{equation*}
Substituting these equations into~(\ref{muSNR}) gives
\begin{align}
\nonumber
\mu_1 & = \gamma \cdot a_1 \cdot \frac{1}{1 - a_1^2} \\
  & = \gamma \cdot \frac{\SNR_1}{a_1}.
\end{align}
Similarly,
\begin{equation}
\mu_2 = \gamma \cdot \frac{\SNR_2}{a_2}.
\end{equation}
Now using the fact that $a_1 a_2 = \rho$, we have
\begin{align}
\nonumber
[a_1 \ \ a_2] \cdot \m{K}_y^{-1} & = \frac{1}{1 - \rho^2} [a_1 - \rho a_2 \ \
    a_2 - \rho a_1] \\
   & = \frac{1}{1 - a_1^2 a_2^2} \left[ \frac{a_1^2 - a_1^2 a_2^2}{a_1} \ \ 
    \frac{a_2^2 - a_1^2 a_2^2}{a_2^2}\right] .
\end{align}
Substituting for $a_1$ and $a_2$ using~(\ref{adef}), this gives
$$
[a_1 \ \ a_2] \cdot \m{K}_y^{-1} = \left[ \frac{\gamma \SNR_1}{a_1} \ \ 
   \frac{\gamma \SNR_2}{a_2} \right].
$$
Thus we have
\begin{equation}
E[x|\ve{y}] = \gamma \left( \SNR_1 \frac{y_1}{a_1} +
  \SNR_2 \frac{y_2}{a_2} \right).
\end{equation}
It follows that
\begin{equation*}
E[x|\ve{y}] = \bmu^T \ve{y},
\end{equation*}
and in particular, $x$ can be written
\begin{equation*}
x = \bmu^T \ve{y} + \tilde{n},
\end{equation*}
where $\tilde{n}$ is Gaussian, independent of $\ve{y}$, and
has variance 
\begin{align}
\nonumber
1 - [a_1 \ \ a_2] \cdot \m{K}_y^{-1} \cdot \left[ \begin{array}{c}
a_1 \\
a_2 
\end{array} \right] 
  & = 
1 - \left[\frac{\gamma \SNR_1}{a_1} \ \ \frac{\gamma \SNR_2}{a_2}\right] 
   \left[ \begin{array}{c}
a_1 \\
a_2
\end{array} \right] \\
  & = \gamma.
\end{align}
Then for any random variable
$u$ such that $x \markov \ve{y} \markov u$, by a calculation
similar to~(\ref{coupleddistortion}), we have
\begin{equation}
\label{coupled}
E[(x - E[x|u])^2] = \gamma + E[(\bmu^T \ve{y} - E[\bmu^T \ve{y}|u])^2].
\end{equation}
As in the proof of Corollary~\ref{remotecor}, it
follows that any code that achieves distortion $d$
for the $\bmu$-sum problem must achieve distortion
$d + \gamma$ for the CEO problem~(\ref{firstcoupling}) 
and vice versa. The characterization of the rate region in~(\ref{muregion})
and the sum rate in~(\ref{convexprob}) 
now follow from existing results on the CEO 
problem~\cite{Oohama:CEO:Region,Prabhakaran:ISIT04}. 
To show that~(\ref{optagain}) equals~(\ref{convexprob}), we first
show that~(\ref{convexprob}) can be rewritten as 
\begin{equation}
\label{rewritten}
\begin{gathered}
\inf\Bigg\{\frac{1}{2}\log
  \Bigg[1 + \sum_{j = 1}^2 \SNR_j (1-2^{-2r_j}) \Bigg] + r_1 + r_2 : 
     r_1 \ge 0, \ r_2 \ge 0, \ \text{and} \\
   1 + \sum_{j = 1}^2 
   \SNR_j \left(1-2^{-2r_j}\right) \ge 
   \frac{1}{d + \gamma}\Bigg\}.
\end{gathered}
\end{equation}
To see this, note that the two optimization problems differ only in
the objective, and both objectives are increasing functions of $r_1$
and $r_2$. Now if $d > 1 - \gamma$, then both infima are zero.
On the other hand, if $d \le 1 - \gamma$, then
in both problems, we may assume without
loss of generality that the constraint is met with equality
\begin{equation}
   1 + \sum_{j = 1}^2 
   \SNR_j \left(1-2^{-2r_j}\right) =
   \frac{1}{d + \gamma}.
\end{equation}
But if the constraint is met with equality, then the two objectives
are equal. Thus the two optimization problems are equivalent.

Let $\ve{u}$ be a distributed Gaussian test channel such that
$$
  x \markov \ve{y} \markov \ve{u}.
$$
If we define
\begin{equation*}
  r_j = I(y_j;u_j|x) \quad j \in \{1,2\},
\end{equation*}
then a standard calculation shows that
\begin{equation}
  E[(x - E[x|\ve{u}])^2] 
   = \left[ 1 + \sum_{j = 1}^2 \SNR_j \left(1 - 2^{-2r_j}\right)\right]^{-1}.
\end{equation}
Thus the expression in~(\ref{rewritten}) equals
\begin{equation}
\label{infauxvar}
\begin{split}
\inf\Big\{I(x;\ve{u}) + I(y_1;u_1|x) + I(y_2;u_2|x)
  : \ & (x,\ve{y},\ve{u}) \ \text{are jointly Gaussian}, \\
   &  x \markov \ve{y} \markov \ve{u} \\
   &  u_1 \markov y_1 \markov y_2 \markov u_2 \\
 & E[(x - E[x|\mathbf{u}])^2] \le d + \gamma\Big\}.
\end{split}
\end{equation}
Now since $(y_1,u_1) \markov x \markov (y_2,u_2)$ and
$x \markov \ve{y} \markov \ve{u}$, we have
\begin{align}
\nonumber
  I(x;\ve{u}) + I(y_1;u_1|x) + I(y_2;u_2|x) & =
  I(x;\ve{u}) + I(\ve{y};\ve{u}|x) \\
\nonumber
              & = I(x,\ve{y};\ve{u}) \\
\label{condind}
             &  = I(\ve{y};\ve{u}).
\end{align}
Applying~(\ref{coupled}) again, we can write the infimum 
in~(\ref{infauxvar}) as
\begin{align*}
\inf\Big\{I(\ve{y};\ve{u}) 
  : \ & (\ve{y},\ve{u}) \ \text{are jointly Gaussian}, \\
   &  u_1 \markov y_1 \markov y_2 \markov u_2 \\
 & E[(\bmu^T \ve{y} - E[\bmu^T \ve{y}|\ve{u}])^2] \le d\Big\},
\end{align*}
which equals
\begin{equation}
\label{optagainrepeat}
   \inf\left\{\frac{1}{2} \log \frac{|\m{K}_y|}{|\m{D}|} : 
    \m{D} \in \mathcal{D}_G \ 
    \text{and}
      \ \bmu^T \m{D} \bmu \le d\right\}.
\end{equation}
Now since the quantity
$$
 1 + \sum_{j = 1}^2 \SNR_j \left(1 - 2^{-2r_j}\right)
$$
is strictly concave, it follows that the infimum in~(\ref{convexprob}) is
achieved by a unique feasible point. This in turn implies that 
the infimum in~(\ref{rewritten}) is achieved by a unique feasible point.
By the equivalence between the feasible points in~(\ref{rewritten})
and~(\ref{optagainrepeat}) it follows that the 
infimum in ~(\ref{optagainrepeat})
is also achieved by a unique feasible point.
\end{proof}

\section{Every $\m{D}^*$ Solves a $\bmu$-sum Problem}
\label{existsmuapp}

\begin{proof}[Proof of Lemma~\ref{existsmu}]
Without loss of generality, we may assume that $\bmu^*$ has been scaled
so that it satisfies the normalization~(\ref{normalized}). Then
the sum rate for the $\bmu^*$-sum problem with allowable 
distortion
\begin{equation*}
d^* := {\bmu^*}^T \m{D}^* \bmu^*
\end{equation*}
is given by
\begin{equation}
\label{convexprob2}
\begin{gathered}
\inf\Bigg\{\frac{1}{2}\log
  \left[\frac{1}{d^* + \gamma}\right] + r_1 + r_2 : 
     r_1 \ge 0, \ r_2 \ge 0, \ \text{and} \\
   1 + \sum_{j = 1}^2 
   \SNR_j \left(1-2^{-2r_j}\right) \ge 
   \frac{1}{d^* + \gamma}\Bigg\},
\end{gathered}
\end{equation}
where $\SNR_1$, $\SNR_2$, and $\gamma$ were defined in
equations~(\ref{SNR1}) through~(\ref{gammadef}). 
The remainder of the proof consists of three parts:
\begin{enumerate}
\item[(\emph{I})] We identify candidate optimizers for~(\ref{convexprob2}), 
  $r_1^*$ and $r_2^*$, in terms of $\m{D}^*$.
\item[(\emph{II})] We show that $r_1^*$ and $r_2^*$ achieve the
  infimum in~(\ref{convexprob2}).
\item[(\emph{III})] We show that at $r_1^*$ and $r_2^*$, the objective
$$
\frac{1}{2} \log \left[\frac{1}{d^* + \gamma}\right] + r_1^* + r_2^*
$$
equals
$$
\frac{1}{2} \log \frac{|\m{K}_y|}{|\m{D}^*|}.
$$
\end{enumerate}

\emph{Part I.} Since $\m{D}^*$ is in $\mathcal{D}_G$, there exists
$\lambda_1^* \ge 0$ and $\lambda_2^* \ge 0$, such that
\begin{equation}
\label{optlambda}
{\m{D}^*}^{-1} = \m{K}_y^{-1} + \left[
\begin{array}{cc}
\lambda_1^* & 0 \\
0 & \lambda_2^*
\end{array}
\right].
\end{equation}
Our candidate optimizers are then
\begin{equation}
\label{rdef}
r^*_j = \frac{1}{2} \log \left(1 + \frac{\lambda_j^*}{1 + \SNR_j}\right)
  \quad j \in \{1,2\}.
\end{equation}
This formula can be understood as follows.
Since $\m{D}^*$ is in
$\mathcal{D}_G$, there exists a distributed Gaussian test
channel $\ve{u}^*$ such that
$\Cov(\ve{y}|\ve{u}^*) = \m{D}^*$. Now $\ve{u}^*$ can be
written
$$
u^*_j = \sqrt{\frac{\lambda_j^*}{1 + \lambda_j^*}} \cdot y_j + z_j
   \quad j \in \{1,2\}
$$
where $\ve{z}$ is an
independent Gaussian vector with covariance matrix
$$
\left[
\begin{array}{cc}
(1 + \lambda_1^*)^{-1} & 0 \\
0 & (1 + \lambda_2^*)^{-1}
\end{array}
\right].
$$
As in the previous appendix, let
\begin{equation}
a_j = \left(\frac{\SNR_j}{1 + \SNR_j}\right)^{1/2} \quad j \in \{1,2\}
\end{equation}
and let $x$, $n_1$, and $n_2$ be zero mean Gaussian random variables with 
\begin{align*}
E[x^2] & = 1 \\
E[n_j^2] & = 1 - a_j^2 \quad j \in \{1,2\}.
\end{align*}
Then couple these variables to $(\ve{y},\ve{u}^*)$ such that
\begin{align*}
y_1 = a_1 x + n_1 \\
y_2 = a_2 x + n_2 
\end{align*}
and $x \markov \ve{y} \markov \ve{u}^*$. It then follows that
\begin{align*}
r^*_j & = \frac{1}{2} \log \left(1 + \lambda_j^*(1 - a_j^2) \right) \\
    & = I(y_j;u_j^*|x).
\end{align*}

\emph{Part II.}
Next we show that $r_1^*$ and $r_2^*$ solve the
optimization problem~(\ref{convexprob2}). Since the optimization problem
is convex, it suffices to show that $r_1^*$ and $r_2^*$ satisfy
the Karush-Kuhn-Tucker (KKT)
conditions~\cite[Section~5.5.3]{Boyd:Convex}.
The Lagrangian for this optimization problem is
\begin{multline}
L(r_1,r_2,\nu) \\
  = \frac{1}{2} \log \left[\frac{1}{d^* + \gamma}\right]
 + r_1 + r_2 - \nu\left(1 + \sum_{j = 1}^2 \SNR_j (1 - 2^{-2 r_j})
   - \frac{1}{d^* + \gamma}\right).
\end{multline}
Thus it suffices to show that
\begin{align}
\label{KKT1}
2^{2 r_j^*} & = \nu^* \SNR_j \quad \forall j \in \{1,2\} \\
\label{KKT2}
 1 + \sum_{j = 1}^2 \SNR_j (1 - 2^{-2r_j^*}) & = \frac{1}{d^* + \gamma}
\end{align}
for some $\nu^* \ge 0$. To show~(\ref{KKT1}), note
that~(\ref{optlambda}) implies that
\begin{equation}
\label{optlambdaD}
\m{D}^* = (1-\rho^2) \left[
\begin{array}{cc}
1 + (1 - \rho^2) \lambda_1^* & -\rho \\
-\rho & 1 + (1- \rho^2)\lambda_2^* 
\end{array} \right]^{-1}.
\end{equation}
Then define
\begin{equation}
\label{sdef}
s_j^* = 1 + (1-\rho^2) \lambda_j^* \quad j \in \{1,2\}.
\end{equation}
Since, by definition,
$$
\m{D}^* = \left[ \begin{array}{cc}
d_1 & \theta^* \sqrt{d_1 d_2} \\
\theta^* \sqrt{d_1 d_2} & d_2
\end{array}
\right],
$$
it follows from~(\ref{optlambdaD}) that
$$
\frac{d_1}{d_2} = \frac{s_2^*}{s_1^*}.
$$
Referring to the definition of $\SNR_1$ and $\SNR_2$, this implies
\begin{align}
\label{sSNR1}
\SNR_1 & = \frac{\rho}{1 - \rho^2} \left( \sqrt{\frac{s_1^*}{s_2^*}}
   + \rho \right) \\
\label{sSNR2}
\SNR_2 & = \frac{\rho}{1 - \rho^2} \left( \sqrt{\frac{s_2^*}{s_1^*}} + 
   \rho \right).
\end{align}
Combining~(\ref{rdef}) and~(\ref{sdef}), we have
$$
s_j^* = (1 - \rho^2) (1 + \SNR_j) (2^{2r_j^*} - 1) + 1.
$$
Thus~(\ref{KKT1}) is equivalent to
\begin{equation}
\label{newKKT1}
\frac{s_j^* - 1}{(1-\rho^2)(1 + \SNR_j)} + 1 = \nu^* \SNR_j
 \quad \forall \ j \in \{1,2\}
\end{equation}
But by using~(\ref{sSNR1}) and~(\ref{sSNR2}),
one can verify that this pair of conditions holds if
\begin{equation}
\nu^* = \frac{1 - \rho^2}{\rho} \frac{s_1^* s_2^* + \rho \sqrt{s_1^* s_2^*}}
    {\rho(s_1^* + s_2^*) + \sqrt{s_1^* s_2^*}(1 + \rho^2)}.
\end{equation}
This establishes~(\ref{newKKT1}) and hence~(\ref{KKT1}). Now
as in the previous appendix we have
\begin{align}
\nonumber
E[(x - E[x|\ve{u}^*])^2] & =
  \left[ 1 + \sum_{j = 1}^2 \SNR_j (1 - 2^{-2r_j^*})\right]^{-1} \\
\nonumber
 & = \gamma + E[({\bmu^*}^T \ve{y} - E[{\bmu^*}^T \ve{y} |\ve{u}^*])^2] \\
\nonumber
 & = \gamma + {\bmu^*}^T \m{D}^* \bmu^* \\
 & = \gamma + d^*.
\end{align}
This establishes~(\ref{KKT2}) and the optimality of
$r_1^*$ and $r_2^*$.

\emph{Part III.}
It only remains to 
show that
\begin{equation}
\frac{1}{2} \log \left[ \frac{1}{d^* + \gamma}\right] + r_1^* + r_2^* =
  \frac{1}{2} \log \frac{|\m{K}_y|}{|\m{D}^*|}.
\end{equation}
Observe that the left-hand side equals
\begin{equation*}
I(x;\ve{u}^*) + \sum_{j = 1}^2 I(y_j;u_j^*|x).
\end{equation*}
Repeating the argument in~(\ref{condind}), we have
\begin{equation*}
I(x;\ve{u}^*) + \sum_{j = 1}^2 I(y_j;u_j^*|x) = I(\ve{y};\ve{u}^*) = 
   \frac{1}{2} \log \frac{|\m{K}_y|}{|\m{D}^*|}. \qedhere
\end{equation*}
\end{proof}

\section{Achievability for $\m{M}$-sums}
\label{Msumsdirectapp}

Before proving Lemma~\ref{Msumsdirect}, we examine some of the properties
of the constituent regions $\mathcal{R}_j^\star(d_j)$.
Without loss of generality, we focus on $\mathcal{R}_1^\star(d_1)$.

Let $h_1(\cdot)$ denote the function whose epigraph is 
$\mathcal{R}_1^\star(d_1)$
\begin{equation}
\mathcal{R}_1^\star(d_1) = \{(R_1,R_2) : R_2 \ge h_1(R_1) \},
\end{equation}
which may equal infinity for some $R_1$. Note that  $h_1(\cdot)$
is nonincreasing. Since $\mathcal{R}_1^\star(d_1)$ is closed and
convex, $h_1(\cdot)$ must be continuous on its effective domain
\cite[Theorems 7.1 and 10.1]{Rockafellar:Convex}. Thus
$h_1(\cdot)$ is closed and proper~\cite{Rockafellar:Convex}.

For $R_1$ in the effective domain of $R_1$, let $\partial h_1(R_1)$
denote the subdifferential of $h_1(\cdot)$ at $R_1$
\begin{equation}
\partial h_1(R_1) = \{s: h_1(R) \ge h_1(R_1) + s(R - R_1) \ \forall \ R\}.
\end{equation}
Note that $\partial h_1(R_1)$ can be interpreted geometrically
as the set of slopes of all supporting lines at $R_1$.
From standard results in convex 
analysis~\cite[Section~24]{Rockafellar:Convex}, $\partial h_1(R_1)$
is the interval between the left and right derivatives of $h_1(\cdot)$
at $R_1$.
The graph of $\partial h_1(\cdot)$,
$$
\{ (R,s) : s \in \partial h_1(R)\}
$$
resembles the graph of a nondecreasing function, except that
any jump discontinuities have been ``filled in'' with vertical segments.
As such, $\partial h_1(\cdot)$ is monotonic in the sense that if
$R_1 \le \tilde{R}_1$  then $s \le \tilde{s}$ for any $s$ in
$\partial h_1(R_1)$ and $\tilde{s}$ in
$\partial h_1(\tilde{R}_1)$.

\begin{figure}
\begin{center}
\scalebox{.65}{\input{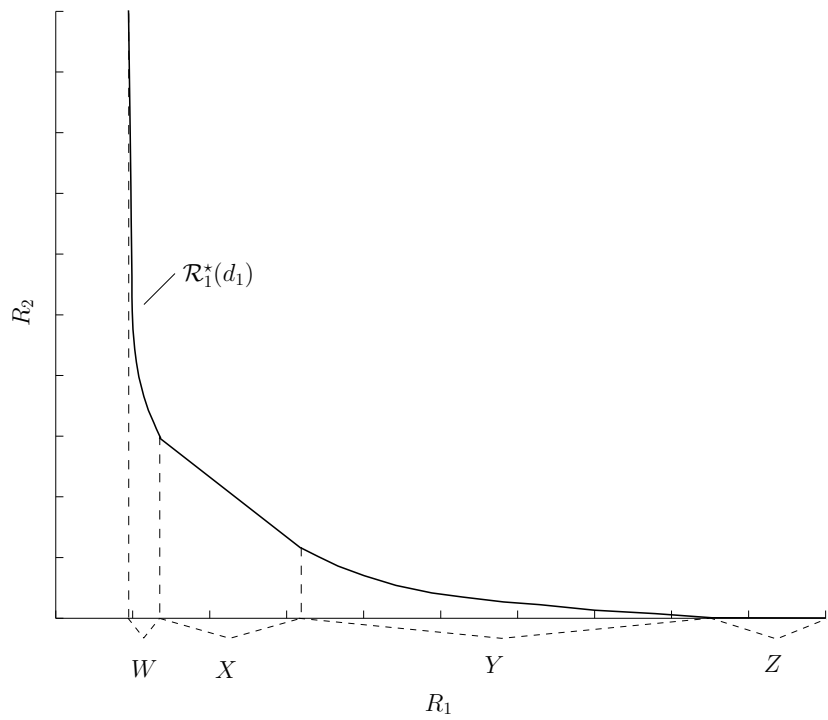}}
\end{center}
\caption{An example illustrating the definitions of 
$W$, $X$, $Y$, and $Z$ in Appendix~\ref{Msumsdirectapp}.}
\label{Rdef}
\end{figure}

We now partition the effective domain of $h_1(\cdot)$ into three
parts. Let
\begin{align}
\nonumber
W & = \{R_1 : \max(\partial h_1(R_1)) < -1\} \\
\nonumber
X & = \{R_1 : \min(\partial h_1(R_1)) \le -1 \le \max(\partial h_1(R_1))\} \\
\nonumber
Y & = \{R_1 : -1 < \min(\partial h_1(R_1)) < 0\} \\
Z & = \{R_1 : \partial h_1(R_1) = \{0\} \}.
\end{align}
These intervals are depicted in Fig.~\ref{Rdef}. We call $W$ the
\emph{steep} part of $\mathcal{R}_1^\star(d_1)$ and $Y$ and $Z$ the
\emph{shallow} part. Some of these intervals might be empty in some cases.
Note that the sum rate decreases as one moves left-to-right in $W$
and increases as one moves left-to-right in $Y$ and $Z$.

Next we associate each point $R_1$ in the effective domain of
$h_1(\cdot)$ with a test channel that meets the $\bmu_1$ distortion
constraint with equality. Since $(R_1,h_1(R_1))$ is on the boundary
of $\mathcal{R}_1^\star(d_1)$, it must be on the boundary of the
contrapolymatroid of some test channel $\m{D}$ satisfying
$\bmu_1^T \m{D} \bmu_1 = d_1.$ Suppose first that $R_1$ is in $W$.
Then $\mathcal{R}_1^\star(d_1)$ has a supporting line with
slope $s < -1$ at $(R_1,h_1(R_1))$. Since the contrapolymatroid
associated with $\m{D}$ is contained in $\mathcal{R}_1^\star(d_1)$,
this contrapolymatroid must also be supported by a line with 
slope $s < -1$ at $(R_1,h_1(R_1))$. This implies that $(R_1,h_1(R_1))$
is on the vertical portion of the boundary of this contrapolymatroid.

In fact, by the definition of $h_1(\cdot)$, $(R_1,h_1(R_1))$ must be
the left corner point of the contrapolymatroid of this test channel.
We then associate $R_1$ with the unique test channel whose left corner
point is $(R_1,h_1(R_1))$. Likewise, to every $R_1$ in $Y$, we associate
the unique test channel whose right corner point is $(R_1,h_1(R_1))$.

If $Z$ is nonempty then $X \cup Y$ is bounded. In this case, as $R_1 
\rightarrow \sup X \cup Y$, the associated test channels will converge
to a test channel $\m{D}$. We associate this test channel with
all $R_1$ in $Z$.

If $R_1$ is in $X$, then $\mathcal{R}_1^\star(d_1)$ is supported
at $(R_1,h_1(R_1))$ by a line with slope $-1$. This implies that
$(R_1,h_1(R_1))$ is sum-rate optimal.
We then associate all
$R_1$ in $X$ with the unique test channel that is sum-rate optimal
(see Lemma~\ref{muCEOextended}).

Note that the end-points of the interval $X$ must correspond to
the corner points of the sum-rate optimal contrapolymatroid.
It follows that the associated test-channels vary continuously with $R_1$
over the entire effective domain of $h_1(\cdot)$.
The test channels also vary monotonically in the sense
that if $R_1 \le \tilde{R}_1$ and $\m{\Lambda}$ and $\tilde{\m{\Lambda}}$
are the associated test channels, then
$$
\m{\Lambda} = \tilde{\m{\Lambda}} + \left[
\begin{array}{cc}
-\lambda_1 & 0 \\
0 & \lambda_2
\end{array}
\right]
$$
for some nonnegative numbers $\lambda_1$ and $\lambda_2$.
Note that for each test channel that meets the distortion constraint
with equality, at least one of its corner points must be on the boundary
of $\mathcal{R}_1^\star(d_1)$.

Now consider a second vector, $\bmu_2$, and suppose that $\bmu_2$
weights $y_2$ more heavily than $\bmu_1$ does
\begin{equation}
\label{weighting}
\frac{\mu_{22}}{\mu_{21}} >
\frac{\mu_{12}}{\mu_{11}}.
\end{equation}
Next we show that as one moves left-to-right along the boundary
of $\mathcal{R}_1^\star(d_1)$, the distortion that the associated
test channels induce on $\bmu^T_2 \ve{y}$ is nondecreasing.

\begin{lemma}
\label{testmonotone}
Suppose $R_1 \le \tilde{R}_1$, and let $\m{D}$
and $\tilde{\m{D}}$ be the associated test channels. Then
\begin{equation}
\label{distdec}
\bmu_2^T \m{D} \bmu_2 \le \bmu_2^T \tilde{\m{D}} \bmu_2.
\end{equation}
\end{lemma}
\begin{proof}
Define $\m{\Lambda}$ and $\tilde{\m{\Lambda}}$ by
\begin{align}
\m{D}^{-1} & = \m{K}^{-1}_y + \m{\Lambda} \\
\tilde{\m{D}}^{-1} & = \m{K}^{-1}_y + \tilde{\m{\Lambda}}.
\end{align}
Since $R_1 \le \tilde{R}_1$, we know that
\begin{equation*}
\m{\Lambda} = \tilde{\m{\Lambda}} +
\left[
\begin{array}{cc}
- \lambda_1 & 0 \\
0 & \lambda_2
\end{array}
\right]
\end{equation*}
for some nonnegative numbers $\lambda_1$ and $\lambda_2$. Then
it follows that
\begin{equation*}
\m{D}^{-1} = \tilde{\m{D}}^{-1} +
\left[ \begin{array}{cc}
- \lambda_1 & 0 \\
0 & \lambda_2
\end{array} \right].
\end{equation*}
Furthermore, we know that
\begin{equation*}
\bmu_1^T \m{D} \bmu_1 = \bmu_1^T \tilde{\m{D}} \bmu_1 = d_1.
\end{equation*}
To establish~(\ref{distdec}), we will show that
\begin{equation}
\label{distdec2}
0 \le \bmu_2^T \tilde{\m{D}} \bmu_2 \bmu_1^T \m{D} \bmu_1
    - \bmu_1^T \tilde{\m{D}} \bmu_1 \bmu_2^T \m{D} \bmu_2.
\end{equation}
Now $\tilde{\m{D}}^{-1}$ can be written
\begin{equation}
\tilde{\m{D}}^{-1} = \left[ \begin{array}{cc}
a & -b \\
-b & c 
\end{array} \right],
\end{equation}
where $a$, $b$, and $c$ are nonnegative. In terms of
these parameters, we have
\begin{align*}
\tilde{\m{D}} & = \frac{1}{a c - b^2} \left[ \begin{array}{cc}
c & b \\
b & a
\end{array} \right] \\
\m{D} & = \frac{1}{(a - \lambda_1)(c+\lambda_2) - b^2} \left[ \begin{array}{cc}
c + \lambda_2 & b \\
b & a-\lambda_1
\end{array} \right] 
\end{align*}
A tedious but straightforward calculation shows that the 
expression on the right-hand
side of~(\ref{distdec2}) equals
\begin{equation*}
\frac{(\lambda_1 c + a\lambda_2) (\mu_{11}^2 \mu_{22}^2 - 
   \mu_{12}^2 \mu_{21}^2) +
  2b (\lambda_1  \mu_{12} \mu_{22} + \lambda_2\mu_{11} \mu_{21})
   (\mu_{11} \mu_{22} -
   \mu_{12} \mu_{21})}{(ac -b^2) ((a - \lambda_1 ) (c + \lambda_2) - b^2)}
\end{equation*}
which is nonnegative because $|\m{D}| > 0$, $|\tilde{\m{D}}| > 0$, and
\begin{equation*}
\mu_{11} \mu_{22} - \mu_{12} \mu_{21} > 0 
\end{equation*}
by~(\ref{weighting}).
\end{proof}
From the proof, we can see that if the inequality in~(\ref{weighting}) is
reversed, then the inequality in~(\ref{distdec}) is also reversed.

We now turn to the proof that
\begin{equation}
\mathcal{R}_\mathrm{sum}^\star (d_1,\ldots,d_J) \cap
  \bigcap_{j = 1}^J \mathcal{R}_j^\star(d_j)
\end{equation}
is achievable. Recall that $\mathcal{R}_\mathrm{sum}^\star(d_1,\ldots
d_J)$ is defined as
$$
\mathcal{R}_\mathrm{sum}^\star(d_1,\ldots,d_J)  =
  \left\{ (R_1,R_2) : R_1 + R_2 \ge R_\mathrm{sum}^\star(d_1,\ldots,d_J)
    \right\}
$$
where
\begin{multline*}
R_\mathrm{sum}^\star(d_1,\ldots,d_J) \\
 = \inf 
  \left\{ \frac{1}{2} \log \frac{|\m{K}_y|}{|\m{D}|} : \m{D} \in \mathcal{D}_G
   \ \text{and} \ \bmu_j^T \m{D} \bmu_j \le d_j \ \forall \ 
    j \in \{1,\ldots,J\} \right\}.
\end{multline*}
We begin by showing that this infimum is achieved.
\begin{lemma}
\label{sumachieved}
There exists a $\m{D}^*$ in $\mathcal{D}_G$ such that
$$
\bmu_j^T \m{D}^* \bmu_j \le d_j \quad \forall j \in \{1,\ldots,J\}
$$
and
$$
\frac{1}{2} \log \frac{|\m{K}_y|}{|\m{D}^*|} =
 \inf \left\{ \frac{1}{2} \log \frac{|\m{K}_y|}{|\m{D}|} : 
   \m{D} \in \mathcal{D}_G
   \ \text{and} \ \bmu_j^T \m{D} \bmu_j \le d_j \ \forall j \in 
     \{1,\ldots,J\}\right\}.
$$
\end{lemma}
\begin{proof}
Let $\tilde{\m{D}}$ by any matrix in $\mathcal{D}_G$ that 
satisfies the distortion constraints. The infimum is then upper
bounded by
$$
\frac{1}{2} \log \frac{|\m{K}_y|}{|\tilde{\m{D}}|}.
$$
Now the set of $\m{D} \in \mathcal{D}_G$ such that
\begin{align}
\bmu_j^T \m{D} \bmu_j & \le d_j \quad \forall j \in \{1,\ldots,J\} \\
\frac{1}{2} \log \frac{|\m{K}_y|}{|\m{D}|} & \le
  \frac{1}{2} \log \frac{|\m{K}_y|}{|\tilde{\m{D}}|},
\end{align}
is compact, and the objective is continuous, so
the infimum is achieved.
\end{proof}

\begin{figure}
\begin{center}
\scalebox{.65}{\input{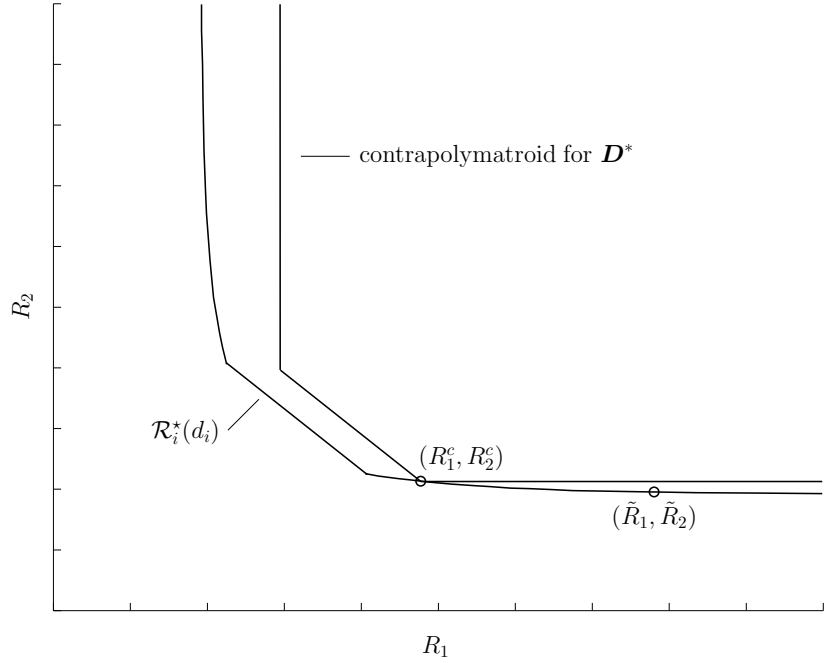}}
\end{center}
\caption{Illustration of $\m{M}$-sums achievability argument.}
\label{Msums:fig}
\end{figure}

\begin{proof}[Proof of Lemma~\ref{Msumsdirect}]
Recall we are assuming that the vectors
 $\bmu_1,\ldots,\bmu_J$ are distinct and have norm one.
We may also assume without loss of generality that they have been ordered
so that their first coordinates are decreasing
\begin{equation*}
\mu_{11} > \mu_{21} > \cdots > \mu_{J1}
\end{equation*}
which implies that their second coordinates  are increasing
\begin{equation*}
\mu_{12} < \mu_{22} < \cdots < \mu_{J2}.
\end{equation*}
Let
\begin{equation}
h_j(R_1) = \inf\{(R_2: (R_1,R_2) \in \mathcal{R}_j^\star(d_j)\}
  \quad j \in \{1,\ldots,J\}
\end{equation}
and
\begin{equation}
h_\mathrm{sum}(R_1) = \inf\{(R_2: (R_1,R_2) \in 
     \mathcal{R}_\mathrm{sum}^\star(d_1,d_2,\ldots,d_J)\}
\end{equation}
denote the functions whose epigraphs are the constituent regions. 
The corresponding function for the intersection is given by
\begin{equation*}
h(R_1) = \max (h_1(R_1),\ldots,h_J(R_1), h_\mathrm{sum}(R_1)).
\end{equation*}
To show that the intersection is achievable, is suffices to
show that for each $R_1$ for which $h(R_1) < \infty$, 
$(R_1,h_j(R_1))$ is achievable for some $j$ or 
$(R_1,h_\mathrm{sum}(R_1))$ is achievable.

By Lemma~\ref{sumachieved},
there exists a $\m{D}^*$ in $\mathcal{D}_G$ that is
sum-rate optimal within this class
\begin{align}
\bmu_j^T \m{D}^* \bmu_j & \le d_j \ \forall j \in \{1,\ldots,J\} \\
\frac{1}{2} \log \frac{|\m{K}_y|}{|\m{D}^*|} & =
  \inf\left\{ \frac{1}{2} \log \frac{|\m{K}_y|}{|\m{D}|} : \m{D} \in 
   \mathcal{D}_G \ \text{and} \
   \bmu_j^T \m{D} \bmu_j \le d_j \ \forall j \in \{1,\ldots, J\}\right\}.
\end{align}
Let $A$ denote the set of constraints that are active
at $\m{D}^*$
\begin{equation*}
A = \{j : \bmu_j^T \m{D}^* \bmu_j = d_j\},
\end{equation*}
which must be nonempty. Let $i$ denote the largest element
of $A$. Thus the $\bmu_i$ sum is the one that weights $y_2$
most heavily of all of the sums whose constraints are active
at $\m{D}^*$.

Consider the contrapolymatroid region associated with $\m{D}^*$.
We shall show that the right corner point of this region is on
the boundary of $\mathcal{R}_i^\star(d_i)$. Since $\m{D}^*$
achieves the $\bmu_i$ constraint with equality, either the
left corner point or the right corner point (or both) of
its contrapolymatroid region must lie on the boundary
of $\mathcal{R}_i^\star(d_i)$. Suppose it is only the left
corner point that lies on the boundary of $\mathcal{R}_i^\star(d_i)$.
This corner point would then have to lie on the steep part of the 
boundary ($W$) of $\mathcal{R}_i^\star(d_i)$. Consider moving a small
amount to the right along the boundary of $\mathcal{R}_i^\star(d_i)$.
Since the distortion for $\bmu_j$ varies continuously as we move
along the boundary of $\mathcal{R}_i^\star(d_i)$ for each $j$,
for a sufficiently small movement, none of the distortion 
constraints in $A^c$ will be violated. At the same time, by
Lemma~\ref{testmonotone}, none of the distortion constraints in $A$ will
be violated either.

Thus we can strictly reduce the sum rate without violating
any of the distortion constraints. This contradicts
the assumption that $\m{D}^*$ is 
sum-rate optimal within the class
$\mathcal{D}_G$. Hence it must be the right corner point,
$(R_1^c,R_2^c)$, of the contrapolymatroid that
is on the boundary of $\mathcal{R}_i^\star(d_i)$, as shown
in Fig.~\ref{Msums:fig}.

Next consider starting at this corner point and moving to the
right along the boundary of $\mathcal{R}_i^\star(d_i)$. By
Lemma~\ref{testmonotone}, the $\bmu_j$ constraint will remain satisfied
for all $j \le i$. If, as we move to the right, the
$\bmu_k$ constraint is never satisfied with equality for
all $k > i$, then it follows that the shallow portion of
$\mathcal{R}_i^\star(d_i)$ to the right of $(R_1^c,R_2^c)$
is achievable.

If the $\bmu_k$ distortion constraint becomes active for some
$k > i$, then there exists a point $(\tilde{R}_1, \tilde{R}_2)$
on the boundary of 
$\mathcal{R}_i^\star(d_i)$ whose associated test 
channel~$\tilde{\m{D}}$ meets
both the $\bmu_i$ and $\bmu_k$ distortion constraints with
equality
\begin{align}
\bmu_i^T \tilde{\m{D}} \bmu_i & = d_i \\
\bmu_k^T \tilde{\m{D}} \bmu_k & = d_k.
\end{align}
Since $(\tilde{R}_1,\tilde{R}_2)$ is on the shallow portion of 
the boundary of $\mathcal{R}_i^\star(d_i)$, $(\tilde{R}_1,\tilde{R}_2)$
must be the right corner point of the contrapolymatroid region associated
with $\tilde{\m{D}}$. Now at least one of the corner points of 
$\tilde{\m{D}}$
must be on the boundary of $\mathcal{R}_k^\star(d_k)$.
By Lemma~\ref{testmonotone},
the boundary of $\mathcal{R}_k^\star(d_k)$ must be below that of
$\mathcal{R}_i^\star(d_i)$ to the left of $\tilde{R}_1$. It follows
that it must be the right corner point of $\tilde{\m{D}}$, i.e.,
$(\tilde{R}_1,\tilde{R}_2)$, that is on the boundary of
$\mathcal{R}_k^\star(d_k)$. We then move to the right along the
boundary of $\mathcal{R}_k^\star(d_k)$, repeating the above process
as necessary as new constraints become active. This
shows that for each $R_1 \ge R_1^c$, $(R_1, h_j(R_1))$ is 
achievable for some $j$. 

An analogous procedure can be followed starting with the left corner
point of the contrapolymatroid associated with $\m{D}^*$.
Finally, between these two corner points, $(R_1, h_\mathrm{sum}(R_1))$
is achievable.
\end{proof}

\section{Converse for $\m{M}$-sums}
\label{Msumsapp}

We show in this appendix that the rate region for the
$\m{M}$-sums problem is contained in
\begin{equation*}
\mathcal{R}_\mathrm{sum}^\star (d_1,\ldots,d_J) \cap
  \bigcap_{j = 1}^J \mathcal{R}_j^\star(d_j).
\end{equation*}
Since the rate region is clearly contained in
$\cap_{j = 1}^J \mathcal{R}_j^\star(d_j)$, we only
need to show that it is contained in 
$\mathcal{R}_\mathrm{sum}^\star (d_1,\ldots,d_J)$.
That is, we must show that the sum rate is lower bounded by
\begin{multline}
R_\mathrm{sum}^\star(d_1,\ldots,d_J) \\
 = \inf 
  \left\{ \frac{1}{2} \log \frac{|\m{K}_y|}{|\m{D}|} : \m{D} \in \mathcal{D}_G
   \ \text{and} \ \bmu_j^T \m{D} \bmu_j \le d_j \ \forall
    j \in \{1,\ldots,J\}\right\}.
\end{multline}
By Lemma~\ref{sumachieved}, this infimum is achieved by some $\m{D}^*$ in 
$\mathcal{D}_G$. If $\m{D}^* = \m{K}_y$, then 
$R^\star_\mathrm{sum}(d_1,\ldots,d_J) = 0$ and our task is trivial.
Suppose therefore that $\m{D}^* \ne \m{K}_y$. Then
$\m{D}^*$ must meet at least one of the
distortion
constraints with equality. If 
$\m{D}^*$ meets exactly one distortion constraint with
equality, then the converse is relatively simple, because
in this case
$\mathcal{R}^\star_\mathrm{sum}(d_1,\ldots,d_J)$
contains 
$\mathcal{R}_j^\star(d_j)$ for some $j$.

\begin{lemma}
\label{oneactive}
Suppose there exists $\m{D}^*$ in $\mathcal{D}_G$ that
is sum-rate optimal within this class and meets exactly
one of the distortion constraints with equality, i.e.,
$\bmu_i^T \m{D}^* \bmu_i = d_i$ and 
$\bmu_j^T \m{D}^* \bmu_j < d_j$ for all $j \ne i$.
Then the sum rate for the $\m{M}$-sums problem is
$$
\frac{1}{2} \log \frac{|\m{K}_y|}{|\m{D}^*|} = 
   R_\mathrm{sum}^\star(d_1,\ldots,d_J).
$$
\end{lemma}
\begin{proof}
It suffices to show that $\m{D}^*$ is sum-rate optimal for the
$\bmu_i$-sum problem with distortion $d_i$.
Recall that one of the corner points of $\m{D}^*$ lies
on the boundary of the $\bmu_i$-sum
rate region. If this point is not sum-rate optimal,
then it is possible to move a small distance along the
boundary of $\mathcal{R}_i^\star(d_i)$  and strictly
decrease the sum rate. Since the distortion associated
with $\bmu_j$ varies continuously as we move along this
boundary for each $j$, it follows that a sufficiently
small movement will not violate any of the distortion
constraints. This contradicts the assumption
that $\m{D}^*$ is sum-rate optimal for the $\m{M}$-sums
problem within the class $\mathcal{D}_G$. It follows that 
$\m{D}^*$ is sum-rate optimal for the $\bmu_i$-sum problem.
\end{proof}

It also happens that 
$\mathcal{R}_\mathrm{sum}^\star(d_1,\ldots,d_J)$ contains
$\mathcal{R}_j^\star(d_j)$ for some $j$ when there is
an optimizing $\m{D}^*$ on the ``boundary'' of $\mathcal{D}_G$.
Let $\mathcal{D}_G^\circ$ denote the set of $\m{D}$ in 
$\mathcal{D}_G$ such that
$$
\m{D}^{-1} = \m{K}_y^{-1} + \m{\Lambda}
$$
for some diagonal and positive definite $\m{\Lambda}$. Let
$\partial \mathcal{D} = \mathcal{D}_G - \mathcal{D}_G^\circ$,
which is the set of those $\m{D}
\in \mathcal{D}_G$ such that $\m{D}^{-1} - \m{K}_y^{-1}$ is
singular.

\begin{lemma}
\label{optboundary}
Suppose there exists a $\m{D}^* \ne \m{K}_y$ 
in $\partial \mathcal{D}_G$ that is
sum-rate optimal within the class $\mathcal{D}_G$.
Then the sum rate for the $\m{M}$-sums problem is given by
$$
\frac{1}{2} \log \frac{|\m{K}_y|}{|\m{D}^*|} = 
   R_\mathrm{sum}^\star(d_1,\ldots,d_J).
$$
\end{lemma}
\begin{proof}
As in the previous proof, it suffices to show that for some $i$,
$\m{D}^*$ is sum-rate optimal for
the $\bmu_i$-sum problem with distortion $d_i$.
Now $\m{D}^*$ can be written
\begin{equation}
\m{D}^* = (\m{K}_y^{-1} + \m{\Lambda})^{-1},
\end{equation}
where $\m{\Lambda}$ is diagonal, positive semidefinite, singular, and
nonzero. Without loss of generality, we may assume that 
$\m{\Lambda}$ is of the form
$$
\m{\Lambda} = \left[ \begin{array}{cc}
\lambda_1 & 0 \\
0 & 0
\end{array} \right]
$$
for some $\lambda_1 > 0$. Let $A$ denote the set of constraints that
are active at $\m{D}^*$
\begin{equation}
A = \{j: \bmu_j^T \m{D}^* \bmu_j = d_j\}.
\end{equation}
Let $i$ denote the constraint in $A$ that weights $y_1$ most heavily
$$
i = \arg \max_{j \in A} \frac{\mu_{j1}}{\mu_{j2}}.
$$
We will show that $\m{D}^*$ is sum-rate optimal for the $\bmu_i$-sum problem
with distortion $d_i$.
Now $\m{D}^*$ is associated with a point
$(R_1,0)$ on the boundary of the $\bmu_i$-sum rate region. If
this point is not sum-rate optimal for this problem, then it is possible to
move a small distance to the left along the boundary of 
$\mathcal{R}_i^\star(d_i)$ and strictly decrease
the sum rate. Since the distortion associated with $\bmu_j$ varies
continuously as one moves along this boundary,
for a sufficiently small movement, none of the constraints in
$A^c$ will be violated. On the other hand, by Lemma~\ref{testmonotone}, none
of the constraints in $A$ will be violated either. This contradicts
the assumption that $\m{D}^*$ is sum-rate optimal for the $\m{M}$-sums
problem within the class $\mathcal{D}_G$. It follows that $\m{D}^*$
is sum-rate optimal for the $\bmu_i$-sum problem.
\end{proof}

The previous two lemmas allow us to focus on the case in which
there exists an optimizing $\m{D}^*$ in $\mathcal{D}_G^\circ$
that meets at least
two of the distortion constraints with equality. Our proof
in this case parallels the proof of Theorem~\ref{main:theorem}. In
particular, we introduce a nonnegative vector $\bmu^*$ such
that $\m{D}^*$ is sum-rate optimal for the $\bmu^*$-sum problem.

\begin{lemma}
\label{existsmuplus}
Suppose $\m{D}^* \in \mathcal{D}_G^\circ$ is sum-rate
optimal within the class $\mathcal{D}_G$
\begin{align}
\nonumber
\bmu_j^T \m{D}^* \bmu_j & \le d_j \quad \forall j \in \{1,\ldots,J\} \\
\label{manysumsumrateopt}
\frac{1}{2} \log \frac{|\m{K}_y|}{|\m{D}^*|} & = \inf \left\{
   \frac{1}{2} \log \frac{|\m{K}_y|}{|\m{D}|} : \m{D} \in \mathcal{D}_G
   \ \text{and} \
   \bmu_j^T \m{D} \bmu_j \le d_j \ \forall j \in \{1,\ldots,J\}\right\}.
\end{align}
Let $A$ denote the set of constraints that are active at $\m{D}^*$
\begin{equation*}
A = \{j : \bmu_j^T \m{D}^* \bmu_j = d_j\}.
\end{equation*}
If $|A| \ge 2$, then there exists $i,j$ in $A$, $i \ne j$, and
a nonnegative vector $\bmu^*$ such that
\begin{enumerate}
\item $\m{D}^*$ is sum-rate optimal for the $\bmu^*$-sum problem,
\begin{equation}
\frac{1}{2} \log \frac{|\m{K}_y|}{|\m{D}^*|} = \inf \left\{
   \frac{1}{2} \log \frac{|\m{K}_y|}{|\m{D}|} : \m{D} \in \mathcal{D}_G
  \ \text{and} \
   {\bmu^*}^T \m{D} \bmu^* \le {\bmu^*}^T \m{D}^* \bmu^* \right\}
\end{equation}
\item $\m{M}_2^{-1} \bmu^*$ is nonnegative, where
\begin{equation}
\m{M}_2 = \left[ \bmu_i \ \ \bmu_j \right].
\end{equation}
\end{enumerate}
\end{lemma}

\begin{proof}
Any $\m{D} \in \mathcal{D}_G$ can be written
\begin{equation*}
\m{D}^{-1} = \m{K}_y^{-1} + \m{\Lambda}
\end{equation*}
for some diagonal and positive semidefinite matrix $\m{\Lambda}$.
Thus the optimization problem in~(\ref{manysumsumrateopt}) can be
written 
\begin{eqnarray}
\min & & \frac{1}{2} \log \frac{|\m{K}_y^{-1} + 
   \m{\Lambda}|}{|\m{K}_y^{-1}|} \\
\text{subject to} & & \bmu_j^T (\m{K}_y^{-1} + 
   \m{\Lambda})^{-1} \bmu_j \le d_j \quad j \in \{1,\ldots,J\},
\end{eqnarray}
where $\m{\Lambda}$ ranges over all diagonal and positive semidefinite
matrices. By hypothesis, this optimization problem is solved by
\begin{equation*}
\m{\Lambda}^* = {\m{D}^*}^{-1} - \m{K}_y^{-1}.
\end{equation*}
Unfortunately, $\m{\Lambda}^*$ may not be 
\emph{regular}~\cite[p.~309]{Bertsekas:Nonlinear}
for this optimization problem. This is an issue because
the KKT conditions may not hold at a local 
minimum that is not regular~\cite[Example~3.1.1]{Bertsekas:Nonlinear}.
We proceed by using a generalization of the KKT conditions
called the \emph{Fritz John 
conditions}~\cite[Sec.~3.3.5]{Bertsekas:Nonlinear}. The
difference between the Fritz John conditions and the KKT
conditions is that the Fritz John conditions include a scalar
for the objective in addition to the multipliers for the constraints.
The Lagrangian for this optimization problem is
\begin{equation}
\frac{1}{2} \log \frac{|\m{K}_y^{-1} + \m{\Lambda}|}{|\m{K}_y^{-1}|} -
     \sum_{j = 1}^J \nu_j (d_j - \bmu_j^T (\m{K}_y^{-1} 
    + \m{\Lambda})^{-1} \bmu_j) - \sum_{j = 1}^2 \nu_{J + j}
   \ve{e}_j^T \m{\Lambda} \ve{e}_j,
\end{equation}
where the second summation handles the constraint that $\m{\Lambda}$ 
must be positive semidefinite. Denote the coordinates of 
$\m{D}^*$ by
\begin{equation}
\m{D}^* = \left[ \begin{array}{cc}
d_1^* & \theta^* \sqrt{d_1^* d_2^*} \\
\theta^* \sqrt{d_1^* d_2^*} & d_2^* 
\end{array} \right].
\end{equation}
By differentiating the Lagrangian with respect to the two diagonal entries
of $\m{\Lambda}$ and using the calculations in Appendix~\ref{derivapp},
we can express the Fritz John 
conditions~\cite[Prop.~3.3.5]{Bertsekas:Nonlinear} as
\begin{align}
\nu_0^* d_1^* & = \sum_{j = 1}^J \nu_j^* (\bmu_j^T \m{D}^* \ve{e}_1)^2 + 
            \nu_{J+1}^* \\
\nu_0^* d_2^* & = \sum_{j = 1}^J \nu_j^* (\bmu_j^T \m{D}^* \ve{e}_2)^2 + 
            \nu_{J+2}^*
\end{align}
for some $\nu_0^*, \ldots, \nu_{J+2}^*$ that are nonnegative
but not all zero. Now since $\m{D}^* \in \mathcal{D}_G^\circ$
by hypothesis,
the constraint that $\m{\Lambda}$ be positive semidefinite is not active
at $\m{\Lambda}^*$. By complimentary slackness, this implies that 
$\nu_{J+1}^* = \nu_{J+2}^* = 0$. Also by complimentary
slackness, $\nu_j^* = 0$ if $j \notin A$, so it follows that
\begin{align}
\label{FJ1}
\nu_0^* d_1^* & = \sum_{j \in A} \nu_j^* (\bmu_j^T \m{D}^* \ve{e}_1)^2 \\
\label{FJ2}
\nu_0^* d_2^* & = \sum_{j \in A} \nu_j^* (\bmu_j^T \m{D}^* \ve{e}_2)^2.
\end{align}
Now if $\nu_0^*$ were equal to zero, then we would have
\begin{equation}
\label{ifnupos}
\sum_{j \in A} \nu_j^* \left[(\bmu_j^T \m{D}^* \ve{e}_1)^2 
   + (\bmu_j^T \m{D}^* \ve{e}_2)^2\right] = 0.
\end{equation}
Since $\bmu_j$ is unit-norm for all $j$ by assumption, 
$(\bmu_j^T \m{D}^* \ve{e}_1)^2
   + (\bmu_j^T \m{D}^* \ve{e}_2)^2$ must be positive for all $j$.
Thus~(\ref{ifnupos}) 
would imply that $\nu_j^* = 0$ for all $j$ in $A$. But this
would contradict the condition that at least one of the dual
variables
$\nu_0^*, \ldots, \nu_{J+2}^*$ is nonzero.
It follows that $\nu_0^*$ is positive, so we can divide through
by $\nu_0^*$ in~(\ref{FJ1}) and~(\ref{FJ2}) to obtain
\begin{align}
d_1^* & = \sum_{j \in A} \frac{\nu_j^*}{\nu_0^*}
                             (\bmu_j^T \m{D}^* \ve{e}_1)^2 \\
d_2^* & = \sum_{j \in A} \frac{\nu_j^*}{\nu_0^*}
                             (\bmu_j^T \m{D}^* \ve{e}_2)^2.
\end{align}
Thus $(d_1^*,d_2^*)$ lies in the convex cone formed by the points
\begin{equation*}
((\bmu_j^T \m{D}^* \ve{e}_1)^2,(\bmu_j^T \m{D}^* \ve{e}_2)^2) \quad j \in A.
\end{equation*}
By Carath\'{e}odory's theorem for convex 
cones~\cite[Corollary~17.1.2]{Rockafellar:Convex}, 
there exists
$i,j$ in $A$, $i \ne j$, such that $(d_1^*,d_2^*)$ lies in the
convex cone of those two points alone
\begin{equation}
\label{KKT3}
\begin{split}
d_1^* & = \tilde{\nu}_i (\bmu_i^T \m{D}^* \ve{e}_1)^2 + 
                  \tilde{\nu}_j (\bmu_j^T \m{D}^* \ve{e}_1)^2 \\
d_2^* & = \tilde{\nu}_i (\bmu_i^T \m{D}^* \ve{e}_2)^2 + 
                    \tilde{\nu}_j (\bmu_j^T \m{D}^* \ve{e}_2)^2.
\end{split}
\end{equation}
Let $\m{M}_2$ denote the 2 x 2 matrix
\begin{equation*}
\m{M}_2 = \left[ \bmu_i \ \ \bmu_j \right].
\end{equation*}
By swapping the roles of $\bmu_i$ and $\bmu_j$ if necessary,
we may assume that $|\m{M}_2| > 0$.

Let
\begin{equation}
\bmu^* = \left[ \begin{array}{c}
\sqrt{d_2^*} \\
\sqrt{d_1^*}
\end{array} \right].
\end{equation}
By Lemma~\ref{existsmu}, $\m{D}^*$ is sum-rate optimal for the 
$\bmu^*$-sum
problem, so it only remains to show that $\m{M}_2^{-1} \bmu^*$ is
nonnegative.

Consider the matrix
\begin{equation}
\tilde{\m{D}} = \left[
\begin{array}{cc}
(\bmu_i^T \m{D}^* \ve{e}_1)^2 & (\bmu_j^T \m{D}^* \ve{e}_1)^2 \\
(\bmu_i^T \m{D}^* \ve{e}_2)^2 & (\bmu_j^T \m{D}^* \ve{e}_2)^2
\end{array}
\right].
\end{equation}
In terms of the components of $\m{D}^*$ and $\m{M}_2$, the determinant
of $\tilde{\m{D}}$ is 
\begin{multline}
|\tilde{\m{D}}| = (\mu_{i1} d_1^* + \mu_{i2} \theta^* \sqrt{d_1^* d_2^*})^2
   (\mu_{j1} \theta^* \sqrt{d_1^* d_2^*} + \mu_{j2} d_2^*)^2 \\
  \mbox{ } - (\mu_{j1} d_1^* + \mu_{j2} \theta^* \sqrt{d_1^* d_2^*})^2
   (\mu_{i1} \theta^* \sqrt{d_1^* d_2^*} + \mu_{i2} d_2^*)^2.
\end{multline}
By expanding the products and using the fact that $|\m{M}_2| > 0$,
one can show that $|\tilde{\m{D}}| > 0$. In particular,
the conditions in~(\ref{KKT3}) can be written as
\begin{align}
\left[
\begin{array}{c}
\tilde{\nu}_i \\
\tilde{\nu}_j
\end{array}
\right]
 & = \tilde{\m{D}}^{-1} 
\left[
\begin{array}{c}
d_1^* \\
d_2^*
\end{array}
\right] \\
  & = \frac{1}{|\tilde{\m{D}}|} \left[
\begin{array}{cc}
(\bmu_j^T \m{D}^* \ve{e}_2)^2 & - (\bmu_j^T \m{D}^* \ve{e}_1)^2 \\
- (\bmu_i^T \m{D}^* \ve{e}_2)^2 & (\bmu_i^T \m{D}^* \ve{e}_1)^2 \\
\end{array}
\right]
\left[
\begin{array}{c}
d_1^* \\
d_2^*
\end{array}
\right].
\end{align}
Since $\tilde{\nu}_i$ and $\tilde{\nu}_j$ are nonnegative, this implies that
\begin{align}
\label{firstcoord}
d_1^* (\bmu_j^T \m{D}^* \ve{e}_2)^2 & \ge d_2^* (\bmu_j^T \m{D}^* \ve{e}_1)^2 \\
\label{secondcoord}
d_2^* (\bmu_i^T \m{D}^* \ve{e}_1)^2 & \ge d_1^* (\bmu_i^T \m{D}^* \ve{e}_2)^2.
\end{align}
Now~(\ref{firstcoord}) can be written
\begin{equation*}
\sqrt{d_1^*} (\mu_{j1} \theta^* \sqrt{d_1^* d_2^*} + \mu_{j2} d_2^*)
  \ge \sqrt{d_2^*} (\mu_{j1} d_1^* + \mu_{j2} \theta^* \sqrt{d_1^* d_2^*}).
\end{equation*}
This can be rearranged to show that
\begin{equation}
\label{firstcoord2}
\mu_{j2} \sqrt{d_2^*} \ge \mu_{j1} \sqrt{d_1^*}.
\end{equation}
Likewise,~(\ref{secondcoord}) implies
\begin{equation}
\label{secondcoord2}
\mu_{i1} \sqrt{d_1^*} \ge \mu_{i2} \sqrt{d_2^*}.
\end{equation}
Now $\m{M}_2^{-1} \bmu^*$ can be written
\begin{align}
\m{M}_2^{-1} \bmu^* & = \frac{1}{\mu_{i1} \mu_{j2} - \mu_{i2} \mu_{j1}}
  \left[ 
\begin{array}{cc}
\mu_{j2} & - \mu_{j1} \\
-\mu_{i2} & \mu_{i1}
\end{array}
\right]
\left[
\begin{array}{c}
\sqrt{d_2^*} \\
\sqrt{d_1^*}
\end{array}
\right] \\
& = \frac{1}{\mu_{i1} \mu_{j2} - \mu_{i2} \mu_{j1}}
\left[
\begin{array}{c}
\mu_{j2} \sqrt{d_2^*} - \mu_{j1} \sqrt{d_1^*} \\
\mu_{i1} \sqrt{d_1^*} - \mu_{i2} \sqrt{d_2^*}
\end{array}
\right]
\end{align}
which is component-wise nonnegative due to~(\ref{firstcoord2}),
(\ref{secondcoord2}), and the nonnegativity of $\det(\m{M}_2)$.
\end{proof}

\begin{lemma}
\label{twoactiveint}
Suppose $(R_1,R_2,d_1,\ldots,d_J)$ is strict-sense achievable and
there exists a $\m{D}^*$ in $\mathcal{D}_G^\circ$ that
is sum-rate optimal within the class $\mathcal{D}_G$ and meets at least
two of the distortion constraints with equality.
Then
$$
R_1 + R_2 \ge \frac{1}{2} \log \frac{|\m{K}_y|}{|\m{D}^*|}
  = R_\mathrm{sum}^\star(d_1,\ldots,d_J).
$$
\end{lemma}

\begin{proof}
We give an abbreviated proof due to the similarity
to the proofs of Lemmas~\ref{central} and~\ref{calc}.
From Lemma~\ref{existsmuplus}, 
there exists two constraint vectors $\bmu_i$ and $\bmu_j$,
$i \ne j$, and a nonnegative vector $\bmu^*$ such that
\begin{enumerate}
\item[(\emph{i})] $\bmu_i^T \m{D}^* \bmu_i = d_i$
\item[(\emph{ii})] $\bmu_j^T \m{D}^* \bmu_j = d_j$
\item[(\emph{iii})] $\m{D}^*$ is sum-rate optimal for the $\bmu^*$-sum 
   problem, and
\item[(\emph{iv})] $\m{M}_2^{-1} \bmu^*$ is nonnegative, where
$$
\m{M}_2 = [ \bmu_i \ \ \bmu_j].
$$
\end{enumerate}
For $\theta$ in $(-1,1)$, let
\begin{equation}
\tilde{\m{D}}_\theta = 
\left[
\begin{array}{cc}
d_i & \theta \sqrt{d_i d_j} \\
\theta \sqrt{d_i d_j} & d_j
\end{array}
\right].
\end{equation}
Also define
\begin{equation}
\tilde{R}_{\mathrm{coop}}(\theta) = \frac{1}{2} \log^+ \frac{|\m{K}_y|
     |\m{M}_2|^2}{|\tilde{\m{D}}_\theta|}
\end{equation}
and
\begin{multline}
\tilde{R}_{\mathrm{sum}}(\theta) \\
  = \inf\Bigg\{\frac{1}{2} \log \frac{|\m{K}_y|}{|\m{D}|} : \m{D} \in
      \mathcal{D}_G \ \text{and} \ {\bmu^*}^T \m{D} \bmu^* \le
          {\bmu^*}^{T} 
  \m{M}_2^{-T} \tilde{\m{D}}_\theta \m{M}_2^{-1} \bmu^* \Bigg\}.
\end{multline}

Now fix some code  that achieves $(R_1,R_2,d_1,\ldots,d_J)$. Let
$\ve{z}^n(i)$ denote the decoder's estimate of $\m{M}_2^T \ve{y}^n(i)$,
which we may assume is the conditional expectation of 
$\m{M}_2^T \ve{y}^n(i)$
given the received messages. Let
\begin{equation}
\tilde{\m{D}} = \frac{1}{n} \sum_{i = 1}^n E\left[(\m{M}_2^T \ve{y}^n(i) - 
   \ve{z}^n(i))(\m{M}_2^T \ve{y}^n(i) - \ve{z}^n(i))^T\right]
\end{equation}
denote the average 
covariance matrix of $\m{M}_2^T \ve{y}^n - \ve{z}^n$. Then
\begin{equation*}
\hat{\m{D}} = \m{M}_2^{-T} \tilde{\m{D}} \m{M}_2^{-1}
\end{equation*}
is the error covariance matrix for the estimate of the 
source, $\ve{y}^n$. As in the proof of Lemma~\ref{central},
$\hat{\m{D}}$ must be positive definite, and the sum rate
of the code must satisfy (c.f.~(\ref{coopboundbefore}))
\begin{align}
R_1 + R_2 & \ge \frac{1}{2} \log^+ \frac{|\m{K}_y|}{|\hat{\m{D}}|} \\
       & = \frac{1}{2} \log^+ \frac{|\m{K}_y||\m{M}_2|^2}{
   |\m{M}_2^T \hat{\m{D}} \m{M}_2|} \\
     & = \frac{1}{2} \log^+ \frac{|\m{K}_y||\m{M}_2|^2}{|\tilde{\m{D}}|}.
\end{align}
In particular, $\tilde{\m{D}}$ must also be
positive definite. Let us write it as
\begin{equation*}
\tilde{\m{D}} = \left[
\begin{array}{cc}
\tilde{d_i} & \tilde{\theta}\sqrt{\tilde{d_i} \tilde{d_j}} \\
\tilde{\theta}\sqrt{\tilde{d_i} 
     \tilde{d_j}} & \tilde{d_j}
\end{array}
\right]
\end{equation*}
where $\tilde{d}_i \le d_i$, $\tilde{d}_j \le d_j$,
and $\tilde{\theta} \in (-1,1)$. Now define 
\begin{equation*}
\phi = \frac{\tilde{\theta} \sqrt{\tilde{d_i} 
     \tilde{d_j}}}{\sqrt{d_i d_j}},
\end{equation*}
and note that $\phi$ is in $(-1,1)$.
Then $\tilde{\m{D}} \preceq \tilde{\m{D}}_\phi$, so it follows that
\begin{equation*}
R_1 + R_2  \ge \frac{1}{2} \log^+ \frac{|\m{K}_y||\m{M}_2|^2}
         {|\tilde{\m{D}}_\phi|} = \tilde{R}_{\mathrm{coop}}(\phi).
\end{equation*}
On the other hand,
\begin{align}
\frac{1}{n} \sum_{i = 1}^n E[({\bmu^*}^T \ve{y}^n(i) - 
     {\bmu^*}^T \m{M}_2^{-T} \ve{z}^n(i))^2] 
        & = {\bmu^*}^T \m{M}_2^{-T} \tilde{\m{D}} \m{M}_2^{-1} \bmu^* \\
	& \le {\bmu^*}^T \m{M}_2^{-T} \tilde{\m{D}}_\phi \m{M}_2^{-1} \bmu^*,
\end{align}
i.e., this code achieves distortion ${\bmu^*}^T \m{M}_2^{-T} 
\tilde{\m{D}}_\phi \m{M}_2^{-1} \bmu^*$ for the $\bmu^*$-sum problem.
Lemma~\ref{muCEO} then implies that
\begin{equation*}
R_1 + R_2 \ge \tilde{R}_{\mathrm{sum}}(\phi).
\end{equation*}
It follows that
\begin{equation}
\label{centralMsums}
R_1 + R_2 \ge \inf_{\theta \in (-1,1)} 
   \max(\tilde{R}_{\mathrm{coop}}(\theta), \tilde{R}_{\mathrm{sum}}(\theta)).
\end{equation}
Now let
\begin{equation}
\label{Dtildestardef}
\tilde{\m{D}}^* = \m{M}_2^T \m{D}^* \m{M}_2,
\end{equation}
which must have diagonal entries $d_i$ and $d_j$, respectively.
Let us write $\tilde{\m{D}}^*$ as 
\begin{equation*}
\tilde{\m{D}}^* = \left[
\begin{array}{cc}
d_i & \tilde{\theta}^* \sqrt{d_i d_j} \\
\tilde{\theta}^* \sqrt{d_i d_j} & d_j
\end{array}
\right].
\end{equation*}
Note that we must have $\tilde{\theta}^* \in (0,1)$ since
$\tilde{\m{D}}^*$ is positive definite and both $\m{M}_2$ and
$\m{D}^*$ have positive entries. For $\theta \ge \tilde{\theta}^*$,
we have
\begin{align}
\max(\tilde{R}_{\mathrm{coop}}(\theta),
\tilde{R}_{\mathrm{sum}}(\theta)) \ge \tilde{R}_{\mathrm{coop}}(\theta) \ge
  \tilde{R}_{\mathrm{coop}}(\tilde{\theta}^*) & = 
    \frac{1}{2} \log^+ \frac{|\m{K}_y||\m{M}_2|^2}
                   {\left|\tilde{\m{D}}^*\right|} \\
    & = \frac{1}{2} \log^+ \frac{|\m{K}_y|}{|\m{D}^*|}.
\end{align}
Since both components of $\m{M}_2^{-1} \bmu^*$ are  nonnegative,
it follows that
\begin{equation*}
{\bmu^*}^T \m{M}_2^{-T} \tilde{\m{D}}_\theta \m{M}_2^{-1} \bmu^*
\end{equation*}
is nondecreasing in $\theta$, which implies 
that $\tilde{R}_{\mathrm{sum}}(\cdot)$
is nonincreasing. Thus if $\theta \le \tilde{\theta}^*$,
\begin{equation*}
\max(\tilde{R}_{\mathrm{coop}}(\theta),
\tilde{R}_{\mathrm{sum}}(\theta)) \ge \tilde{R}_{\mathrm{sum}}(\theta) \ge
  \tilde{R}_{\mathrm{sum}}(\tilde{\theta}^*).
\end{equation*}
But observe that at $\tilde{\theta}^*$, $\tilde{R}_\mathrm{sum}(\cdot)$
satisfies
\begin{equation}
\tilde{R}_{\mathrm{sum}}(\tilde{\theta}^*) = 
   \inf\Bigg\{\frac{1}{2} \log \frac{|\m{K}_y|}{|\m{D}|}
  : \m{D} \in \mathcal{D}_G \ \text{and} \ {\bmu^*}^T \m{D} \bmu^* \le
     {\bmu^*}^T \m{M}_2^{-1} \tilde{\m{D}}^* \m{M}_2^{-1} \bmu^* \Bigg\}
\end{equation}
and~(\ref{Dtildestardef}) implies that
\begin{equation*}
\m{M}_2^{-T} \tilde{\m{D}}^* \m{M}_2^{-1} = \m{D}^*.
\end{equation*}
Since $\m{D}^*$ solves the $\bmu^*$-sum problem, this implies
\begin{equation*}
\tilde{R}_{\mathrm{sum}}(\tilde{\theta}^*) = 
     \frac{1}{2} \log \frac{|\m{K}_y|}
      {|\m{D}^*|}.
\end{equation*}
It follows that
\begin{equation}
\inf_{\theta \in (-1,1)} \max(\tilde{R}_{\mathrm{coop}}(\theta),
  \tilde{R}_{\mathrm{sum}}(\theta)) = \frac{1}{2} \log \frac{|\m{K}_y|}
       {|\m{D}^*|}.
\end{equation}
Combining this with~(\ref{centralMsums}) yields the desired conclusion.
\end{proof}

We are now in a position to complete the proof of Theorem~\ref{Msums:theorem}.

\begin{proof}[Proof of Theorem~\ref{Msums:theorem}]
We only need to show that the rate region is contained in
$\mathcal{R}_\mathrm{sum}^\star(d_1,\ldots,d_J)$.
Lemmas~\ref{oneactive}, \ref{optboundary}, 
and \ref{twoactiveint} together imply that if
$$(R_1,R_2,d_1,\ldots,d_J)$$ is strict-sense achievable, then
\begin{equation}
\label{Msumsbigfinish}
R_1 + R_2
  \ge \inf \left\{ \frac{1}{2} \log \frac{|\m{K}_y|}{|\m{D}|} :
   \m{D} \in \mathcal{D}_G \ \text{and} \ \bmu_i^T \m{D} \bmu_i \le d_j
  \ \forall j \in \{1,\ldots,J\} \right\}.
\end{equation}
It is readily verified that the right-hand side is lower-semicontinuous
in $d_1,\ldots,d_J$. It follows that~(\ref{Msumsbigfinish}) also holds
if $(R_1,R_2,d_1,\ldots,d_J)$ are achievable. This implies the desired
conclusion.
\end{proof}

\section{Converse for Many Sources}
\label{manyapp}

In this appendix we prove Theorem~\ref{manysource:theorem}. 
Recall we are
assuming that the covariance matrix $\m{K}_y$ has the form
$$
\m{K}_y = \left[
\begin{array}{cccc}
1 & \rho & \cdots & \rho \\
\rho & 1 & \ldots & \rho \\
\vdots &  \vdots & \ddots & \vdots \\
\rho & \rho & \cdots & 1
\end{array}
\right].
$$
for some $0 < \rho < 1$ and all of the distortion constraints are 
equal to $d$. Our goal is to show that the minimum sum rate, 
$R^\star_\mathrm{sum}(d)$, equals
\begin{equation}
\inf 
  \left\{ \frac{1}{2} \log \frac{|\m{K}_y|}{|\m{D}|} : \m{D} \in \mathcal{D}_G
  \ \text{and} \ \ve{e}_\ell^T \m{D} \ve{e}_\ell \le d \ \forall \
   \ell \in \{1,\ldots,L\} \right\},
\end{equation}
and that this infimum is achieved by a $\m{D} \in \mathcal{D}_G$ of the
form
\begin{equation}
\label{idtest}
\m{D}^{-1} = \m{K}_y^{-1} + \lambda \m{I}
\end{equation}
for some $\lambda \ge 0$. The conclusion 
is obvious if $d \ge 1$, so we will assume
that $0 < d < 1$. Using the matrix inversion 
lemma~\cite[p.~50]{Golub:VanLoan}, the matrix
inversions in~(\ref{idtest}) can be computed explicitly
\begin{equation}
\label{Dexplicit}
\m{D} = \frac{1 - \rho}{1 + \lambda(1- \rho)} \left[ \m{I} +
  \frac{\rho \ve{1} \ve{1}^T}{
   1 - \rho + \lambda(1-\rho)(1 - \rho + \rho L)} \right].
\end{equation}
It follows that there is a unique $\m{D} \in \mathcal{D}_G$ 
of the form in (\ref{idtest}) 
with all of the diagonal entries equal to $d$. Let us call
this matrix $\m{D}^*$. Note that $\m{D}^*$ must be of the form
\begin{equation}
\label{thetastardef}
\m{D}^* = d \left[
\begin{array}{cccc}
1 & \theta^* & \cdots & \theta^* \\
\theta^* & 1 & \ldots & \theta^* \\
\vdots &  \vdots & \ddots & \vdots \\
\theta^* & \theta^* & \cdots & 1
\end{array}
\right]
\end{equation}
for some $\theta^* > 0$. Since $\m{D}^*$ must be positive definite,
it follows that $\theta^* < 1$. Now the inequalities
\begin{align}
R^\star_\mathrm{sum}(d) & \le \inf \left\{ \frac{1}{2} \log 
   \frac{|\m{K}_y|}{|\m{D}|} : \m{D} \in \mathcal{D}_G
  \ \text{and} \ \ve{e}_\ell^T \m{D} \ve{e}_\ell \le d \ \forall
  \ell \in \{1,\ldots,L\}\right\} \\
   & \le \frac{1}{2} \log \frac{|\m{K}_y|}{|\m{D}^*|}
\end{align}
are clear, so it suffices to show that
$$
\frac{1}{2} \log \frac{|\m{K}_y|}{|\m{D}^*|} \le R^\star_\mathrm{sum}(d).
$$

For $\theta$ in $(-1/(L-1),1)$ let
\begin{equation*}
\m{D}_\theta = d \left[
\begin{array}{cccc}
1 & \theta & \cdots & \theta \\
\theta & 1 & \ldots & \theta \\
\vdots &  \vdots & \ddots & \vdots \\
\theta & \theta & \cdots & 1
\end{array}
\right]
\end{equation*}
and note that $\m{D}_\theta$ is positive definite for each $\theta$.
Then define
\begin{equation}
R_{\mathrm{coop}}(\theta) = \frac{1}{2} \log^+ \frac{|\m{K}_y|}{|\m{D}_\theta|}
   = \frac{1}{2} \log^+ \frac{|\m{K}_y|}{d^L (1-\theta)^{L-1}(1 - \theta
  + L\theta)}.
\end{equation}

Next consider the problem of reproducing the sum of the sources,
$\ve{1}^T \ve{y}$, at the decoder. By following the proof of 
Lemma~\ref{muCEOextended}, one can show that this problem is equivalent
to the CEO problem
\begin{equation*}
y_\ell = x + n_\ell \quad \ell \in \{1,\ldots,L\}
\end{equation*}
where $x, n_1, \ldots, n_L$ are zero-mean, Gaussian, and independent,
and $x$ has variance $\rho$. It follows from existing
results that the separation-based
scheme
achieves the entire rate region, and in
particular it is sum-rate optimal~\cite{Oohama:CEO:Region,Prabhakaran:ISIT04}.
Thus 
the sum rate for the $\ve{1}$-sum problem with distortion constraint
$\ve{1}^T \m{D}_\theta \ve{1}$ is given by
\begin{equation}
\label{one-sum-sum}
R_{\mathrm{sum}}(\theta) := \inf\left\{\frac{1}{2} \log \frac{|\m{K}_y|}
    {|\m{D}|} : \m{D} \in \mathcal{D}_G \ \text{and} \ \ve{1}^T \m{D} \ve{1}
  \le \ve{1}^T \m{D}_\theta \ve{1} \right\}.
\end{equation}
In fact, the CEO results imply that 
the infimum in~(\ref{one-sum-sum}) will
be achieved by a $\m{D} \in \mathcal{D}_G$ of the form in~(\ref{idtest}).
In particular, we have
\begin{equation}
\label{staroptstar}
R_\mathrm{sum}(\theta^*) = 
\frac{1}{2} \log \frac{|\m{K}_y|}{|\m{D}^*|}.
\end{equation}

\begin{lemma}
\label{manysource:bound}
If $(R_1,\ldots,R_L,d,\ldots,d)$ is strict-sense achievable, then
\begin{equation}
\label{manysource:inf}
\sum_{\ell = 1}^L R_\ell \ge \inf_{\theta \in (-1/(L-1),1)} \max
  (R_\mathrm{coop}(\theta),R_\mathrm{sum}(\theta)).
\end{equation}
\end{lemma}

\begin{proof}
By hypothesis, there exists a code 
$(f_1^{(n)},\ldots,f_L^{(n)},\varphi_1^{(n)},\ldots,\varphi_L^{(n)})$
such that
\begin{align*}
R_\ell & \ge \frac{1}{n} \log M_\ell^{(n)} \quad \text{for all $\ell$ in 
            $\{1,\ldots,L\}$} \\
d & \ge \frac{1}{n} \sum_{i = 1}^n E[(y_\ell^n(i) - \hat{y}_\ell^n(i))^2]
  \quad \text{for all $\ell$ in $\{1,\ldots,L\}$}.
\end{align*}
From this code, we construct a new code with block length $N := L! \cdot n$
by time-sharing among all $L!$ permutations of the sources.
For this new code, the rates are symmetric
\begin{align*}
\frac{1}{L} \sum_{i = 1}^L R_i & \ge \frac{1}{N}
  \log M_{\ell}^{(N)} \quad \text{for all $\ell$ in $\{1,\ldots,L\}$} \\
\intertext{and the distortion still satisfies}
d & \ge \frac{1}{N} \sum_{i = 1}^N E[(y_\ell^N(i) - \hat{y}_\ell^N(i))^2]
   \quad \text{for all $\ell$ in $\{1,\ldots,L\}$}.
\end{align*}
Let
\begin{equation*}
\hat{\m{D}} = \frac{1}{N} \sum_{i = 1}^N E[(\ve{y}^N(i) - \hat{\ve{y}}^N(i))
   (\ve{y}^N(i) - \hat{\ve{y}}^N(i))^T]
\end{equation*}
denote the error covariance matrix of the code. By the symmetry of the
time sharing, $\hat{\m{D}}$ must have the form
\begin{equation*}
\hat{\m{D}} = \hat{d} \left[
\begin{array}{cccc}
1 & \phi & \cdots & \phi \\
\phi & 1 & \ldots & \phi \\
\vdots &  \vdots & \ddots & \vdots \\
\phi & \phi & \cdots & 1
\end{array}
\right]
\end{equation*}
for some $\hat{d} \le d$. Following the calculation at the 
beginning of the proof of Lemma~\ref{central}, one can show that
$\hat{\m{D}}$ is positive definite, which implies that 
$-1/(L-1) < \phi < 1$, and
$$
\sum_{\ell = 1}^L R_\ell \ge \frac{1}{2} \log^+ 
   \frac{|\m{K}_y|}{|\hat{\m{D}}|}.
$$
But $\hat{\m{D}} \preceq \m{D}_\phi$, so this implies
\begin{equation}
\label{manysource:first}
\sum_{\ell = 1}^L R_\ell \ge \frac{1}{2} \log \frac{|\m{K}_y|}{|\m{D}_\phi|}
   = R_\mathrm{coop}(\phi).
\end{equation}
Since the code has error covariance matrix $\hat{\m{D}}$, the distortion
it achieves for the $\ve{1}$-sum problem is at most 
$\ve{1}^T \hat{\m{D}} \ve{1} \le \ve{1}^T \m{D}_\phi \ve{1}$.
It follows that
$$
\sum_{\ell = 1}^L R_\ell \ge R_\mathrm{sum}(\phi).
$$
Combining this with~(\ref{manysource:first}) gives
$$
\sum_{\ell = 1}^L R_\ell \ge \max(R_\mathrm{sum}(\phi),R_\mathrm{coop}(\phi)).
$$
The conclusion follows by taking the infimum over $\phi$ in $(-1/(L-1),1)$
\end{proof}

Next we evaluate the infimum in~(\ref{manysource:inf}). Recall
that $\theta^*$ is defined by~(\ref{thetastardef}). 

\begin{lemma}
\label{manysource:eval}
$$
\inf_{\theta \in (-1/(L-1),1)} 
  \max (R_\mathrm{coop}(\theta),R_\mathrm{sum}(\theta))
  = \frac{1}{2} \log \frac{|\m{K}_y|}{|\m{D}^*|}.
$$
\end{lemma}

\begin{proof}
By differentiating, one can verify that $R_\mathrm{coop}(\cdot)$ is
nondecreasing on $(0,1)$. Then if $\theta \ge \theta^*$, we have
\begin{equation}
\max(R_\mathrm{coop}(\theta),R_\mathrm{sum}(\theta)) \ge 
  R_\mathrm{coop}(\theta) \ge R_\mathrm{coop}(\theta^*)
    = \frac{1}{2} \log \frac{|\m{K}_y|}{|\m{D}^*|}.
\end{equation}
On the other hand, if $\theta \le \theta^*$, then
since $R_\mathrm{sum}(\cdot)$ is nonincreasing
\begin{equation}
\max(R_\mathrm{coop}(\theta),R_\mathrm{sum}(\theta)) \ge 
  R_\mathrm{sum}(\theta) \ge R_\mathrm{sum}(\theta^*)
    = \frac{1}{2} \log \frac{|\m{K}_y|}{|\m{D}^*|},
\end{equation}
where we have used~(\ref{staroptstar}).
\end{proof}
Theorem~\ref{manysource:theorem} 
now follows from a continuity
argument similar to the one used in the proof of 
Theorem~\ref{main:theorem}.

\section{Derivatives}
\label{derivapp}

Let $\m{D}(\lambda_1,\lambda_2)$ denote the matrix in $\mathcal{D}_G$
defined by
\begin{equation}
\m{D}(\lambda_1,\lambda_2) = \left(\m{K}_y^{-1} + \left[
\begin{array}{cc}
\lambda_1 & 0 \\
0 & \lambda_2
\end{array}
\right]\right)^{-1}.
\end{equation}
In this appendix we compute the derivatives of 
$$\log |\m{D}(\lambda_1,\lambda_2)|$$ and 
$$\bmu^T \m{D}(\lambda_1,\lambda_2) \bmu$$
with respect to $\lambda_1$ and $\lambda_2$. Write
\begin{equation*}
\m{\Lambda} = \left[
\begin{array}{cc}
\lambda_1 & 0 \\
0 & \lambda_2
\end{array}
\right].
\end{equation*}
Then we have
\begin{align}
\frac{\partial \log |\m{D}(\lambda_1,\lambda_2)|}{\partial\lambda_1} & =
   - \lim_{\delta \rightarrow 0} 
      \frac{\log|\m{K}_y^{-1} + \m{\Lambda} + \delta \ve{e}_1 \ve{e}_1^T| -
           \log|\m{K}_y^{-1} + \m{\Lambda}|}{\delta} \\
      & = - \lim_{\delta \rightarrow 0} \frac{\log |\m{I} + \delta
	   \ve{e}_1 \ve{e}_1^T \m{D}(\lambda_1,\lambda_2)|}{\delta} \\
      & = - \lim_{\delta \rightarrow 0} \frac{\log (1 + \delta
	   \ve{e}_1^T \m{D}(\lambda_1,\lambda_2) \ve{e}_1)}{\delta} \\
      & = - \log(e) \ve{e}_1^T \m{D}(\lambda_1,\lambda_2) \ve{e}_1.
\end{align}
Similarly,
\begin{equation}
\frac{\partial \log |\m{D}(\lambda_1,\lambda_2)|}
   {\partial\lambda_2} =  - \log(e) 
      \ve{e}_2^T \m{D}(\lambda_1,\lambda_2) \ve{e}_2.
\end{equation}
Likewise, we have
\begin{align}
& \frac{\partial \bmu^T \m{D}(\lambda_1,\lambda_2) \bmu}{\partial \lambda_1} \\
  & = \lim_{\delta \rightarrow 0} \frac{\bmu^T \left( \left(
     \m{K}_y^{-1} + \m{\Lambda} + \delta \ve{e}_1 \ve{e}_1^T\right)^{-1}
       - (\m{K}_y^{-1} + \m{\Lambda})^{-1}\right) \bmu}{\delta} \\
  & = - \lim_{\delta \rightarrow 0} \bmu^T \left( 
         \m{D}(\lambda_1,\lambda_2) \ve{e}_1 (1 + \delta
	    \ve{e}_1^T \m{D}(\lambda_1,\lambda_2) \ve{e}_1)^{-1}
	      \ve{e}_1^T \m{D}(\lambda_1,\lambda_2)\right)\bmu \\
  & = - (\bmu^T \m{D}(\lambda_1,\lambda_2) \ve{e}_1)^2
\end{align}
where to obtain the second equation we have used the matrix inversion
lemma. Similarly,
\begin{equation}
\frac{\partial \bmu^T \m{D}(\lambda_1,\lambda_2) \bmu}{\partial \lambda_2} =
   -(\bmu^T \m{D}(\lambda_1,\lambda_2) \ve{e}_2)^2.
\end{equation}

\bibliography{awagnerabrv,IEEEabrv,bib}
\bibliographystyle{IEEEtran}

\newpage

\end{document}